\documentclass[aps,psfig,epsfig,preprint] {revtex4}
\usepackage{epsfig}
\usepackage{color}
\usepackage{bm}
\usepackage{subfigure}

\begin{document}
\draft
\title{
{\bf Fermion Mass Hierarchies and Flavor Mixing from $T'$ Symmetry
}}

\author{Gui-Jun Ding}

\affiliation{\centerline{Department of Modern
Physics,}\centerline{University of Science and Technology of
China, Hefei, Anhui 230026, China}} 

\begin{abstract}
We construct a supersymmetric model based on $T'\otimes Z_3\otimes
Z_9$ flavor symmetry. At the leading order, the charged lepton mass
matrix is not diagonal, $T'$ is broken completely, and the hierarchy
in the charged lepton masses is generated naturally. Nearly
tri-bimaximal mixing is predicted, subleading effects induce
corrections of order $\lambda^2$, where $\lambda$ is the Cabibbo
angle. Both the up quark and down quark mass matrices textures of
the well-known $U(2)$ flavor theory are produced at the leading
order, realistic hierarchies in quark masses and CKM matrix elements
are obtained. The vacuum alignment and subleading corrections are
discussed in detail.

\vskip 0.5cm


\end{abstract}
\maketitle
\section{introduction}
Experimental data on the quark and lepton masses and mixing provide
important clues to the nature of new physics beyond the Standard
Model(SM). However, in SM the Yukawa coupling constants which are
responsible for the fermion masses and mixing, can be freely
adjusted without disturbing the internal consistency of the theory,
one must rely on experiments to fix their values. The origin of
fermion mass hierarchies and flavor mixing is a longstanding puzzle
in the SM of particle physics.

Family symmetry is a fascinating idea to this issue. Current data
strongly suggests that there should be a new symmetry that acts
horizontally across the three standard model family\cite{review}.
Ideally, only the top quark Yukawa coupling is allowed by this
symmetry, and all the remaining couplings are generated, as this
symmetry is spontaneously broken down. In the original work of
Froggatt and Nielsen, they suggested the continuous Abelian $U(1)$
as the flavor symmetry, its spontaneous breaking produces the
correct orders of quark mass hierarchies and
Cabibbo-Kobayashi-Maskawa (CKM) matrix
elements\cite{Froggatt:1978nt}. Models with various horizontal
symmetries gauged or global, continuous or discrete, Abelian or
non-Abelian, have been proposed\cite{horizontal}. Recently, it is
found that discrete group $A_4$ is especially suitable to derive the
so-called tri-bimaximal(TB) mixing\cite{TBmix} in the lepton sector
in a natural
way\cite{Ma:2001dn,Babu:2002dz,Ma:2004zv,Altarelli:2005yp,Chen:2005jm,Zee:2005ut,Altarelli:2005yx,He:2006dk,Ma:2006sk,King:2006np,
Morisi:2007ft,Bazzocchi:2007na,Bazzocchi:2007au,Lavoura:2007dw,Brahmachari:2008fn,Altarelli:2008bg,Bazzocchi:2008rz}.
The left-handed electroweak lepton doublets $l_i(i=1,2,3)$ transform
as $A_4$ triplet, the right-handed charged leptons $e^{c}$,
$\mu^{c}$ and $\tau^{c}$ transform as $\mathbf{1}$, $\mathbf{1}''$
and $\mathbf{1}'$ respectively, and two triplets $\varphi_{T}$ and
$\varphi_{S}$ and a singlet $\xi$ are introduced to break the $A_4$
symmetry spontaneously\cite{Altarelli:2005yx}. If we adopt for quark
the same classification scheme under $A_{4}$ that we have used for
leptons, an identity CKM mixing matrix is obtained at the leading
order, which is a good first order approximation. The non leading
corrections in the up and down quark sector almost exactly cancel in
the mixing matrix. It seems very difficult to implement $A_4$ as a
family symmetry for both the quark and lepton sectors.

Double tetrahedral group $T'$ has three inequivalent irreducible
doublet representations $\mathbf{2}$, $\mathbf{2}'$, $\mathbf{2}''$
in addition to the triplet representation $\mathbf{3}$ and three
singlet representations $\mathbf{1}$, $\mathbf{1}'$, $\mathbf{1}''$
as $A_4$. Furthermore, the kronecker products of the triplet and
singlet representations are identical to those of $A_4$. Therefore
$T'$ can reproduce the success of $A_4$ model building in the lepton
sector, and $T'$ as a family symmetry for both quark and lepton has
been
considered\cite{Carr:2007qw,Feruglio:2007uu,Chen:2007afa,Frampton:2007et,Aranda:2007dp,Frampton:2007gg,Frampton:2007hs}.
In Ref.\cite{Feruglio:2007uu} a supersymmetric (SUSY) model with
$T'\otimes Z_3\otimes U(1)_{FN}$ flavor symmetry is presented, which
is identical to $A_4$ in the lepton sector. While the quark doublet
and the antiquarks of the third generations transforms as
$\mathbf{1}$ under $T'$, the other quark doublets and the antiquarks
transforms as $\mathbf{2}''$. TB mixing is derived naturally as in
$A_4$ model, whereas only the masses of the second and the third
generation quarks and the mixing between them are generated at the
leading order. The masses and mixing angles of the first generation
quark are induced by higher dimensional operators. The authors built
a model with $T'\otimes Z_{12}\otimes Z_{12}$ flavor symmetry in the
context of SU(5) grand unification in Ref.\cite{Chen:2007afa}. Both
the quarks and leptons are assumed to transform as
$\mathbf{2}\oplus\mathbf{1}$ under $T'$ in Ref.\cite{Aranda:2007dp}.
A renormalizable model with $T'\otimes Z_2\otimes Z'_2\otimes Z''_2$
flavor symmetry is presented in Ref.\cite{Frampton:2007hs}, where
the flavor symmetry breaking scale is very low in the range 1 GeV-10
GeV.

The $T'$ symmetry can replicate the success of $A_4$ model, and it
allows the heavy third family to to be treated differently,
therefore $T'$ is a very promising flavor symmetry to understand the
origin of fermion mass hierarchies and flavor mixing. In this work
we shall build a SUSY model based on the $T'\otimes Z_3 \otimes Z_9$
flavor symmetry, the transformation rules of $l_i$, $e^{c}$,
$\mu^{c}$ and $\tau^{c}$ are the same as those in the $A_4$
model\cite{Altarelli:2005yx}. In the quark sector, we exploit the
singlet and doublet representation. The fermion mass hierarchies are
generated via the spontaneous breaking of the discrete flavor
symmetry in contrast with
Ref.\cite{Altarelli:2005yx,Feruglio:2007uu}. The Yukawa matrices of
the up and down quarks have the same textures as those in the
well-known U(2) flavor theory\cite{u2f}. The hierarchies in the
masses of the known quarks and leptons, the realistic pattern of CKM
matrix elements and the TB mixing are naturally produced.

The paper is organized as follows. In section II we present the
current experimental data and the parameterizations of fermion mass
hierarchies and flavor mixing. A short review of model with U(2)
flavor symmetry is given in section III; In section IV a model with
$T'\otimes Z_3\otimes Z_9$ flavor symmetry is constructed, its basic
features and predictions are discussed. We present the vacuum
alignment and the subleading corrections to the leading order
results in section V and section VI respectively. We summarize our
results in section VII. Appendix A gives the basic properties of the
$T'$ group. The corrections to the vacuum alignment induced by
higher dimensional operators are discussed in Appendix B.

\section{current experimental data on fermion mass hierarchies and flavor mixing and their parameterizations }

The observed fermion mass hierarchy is apparent in the quark sector.
The masses of up type quarks are\cite{pdg}
\begin{eqnarray}
\nonumber m_{u}&\simeq& 1.5-3 \rm{MeV}\\
\nonumber m_{c}&\simeq&  1.16-1.34 \rm{GeV}\\
\label{1} m_{t}&\simeq& 170.9.1-177.5 \rm{GeV}
\end{eqnarray}
and the masses of down type quarks are
\begin{eqnarray}
\nonumber m_{d}&\simeq& 3-7\rm{MeV}\\
\nonumber m_{s}&\simeq& 70-120 \rm{MeV}\\
\label{2} m_{b}&\simeq& 4.13-4.27 \rm{GeV}
\end{eqnarray}
We note that all the quark masses except the top quark mass are
given in the $\overline{\rm{MS}}$ scheme. The light $u$, $d$, $s$
quark masses are estimates of so-called current quark mass at the
scale about 2 GeV. There is some ambiguity in the measurement of the
absolute quark masses since they are scheme dependent, but the
ratios of the masses are more concrete
\begin{eqnarray}
\nonumber&& \frac
{m_{u}}{m_{d}}\simeq 0.3-0.6\\
\nonumber&& \frac{m_{s}}{m_{d}}\simeq 17-22\\
\label{3}&&\frac{m_{s}-(m_{u}+m_{d})/2}{m_{d}-m_{u}}\simeq 30-50
\end{eqnarray}
The masses of the charged leptons have been measured much more
unambiguously than the quark masses. The charged lepton sector is
also seen to exhibit a large mass hierarchy. Their masses are
measured to be\cite{pdg}
\begin{eqnarray}
\nonumber m_{e}&\simeq& 0.511 \rm{MeV}\\
\nonumber m_{\mu}&\simeq& 105.7 \rm{MeV}\\
\label{4} m_{\tau}&\simeq& 1777 \rm{MeV}
\end{eqnarray}
The $e$, $\mu$ and $\tau$ masses are the pole masses, and their mass
hierarchy is similar to that in the down type quark sector.
Including the renormalization group equation evolution, the fermion
mass ratios at the GUT scale are parameterized in terms of the
Cabibbo angle $\lambda\simeq 0.23$ as
follows\cite{Fusaoka:1998vc,Ross:2007az,Xing:2007fb}
\begin{eqnarray}
\nonumber &&\frac{m_{u}}{m_{t}}\sim
\lambda^{8},~~~~~~~~\frac{m_{c}}{m_{t}}\sim\lambda^{4},\\
\nonumber &&\frac{m_{d}}{m_{b}}\sim
\lambda^{4},~~~~~~~~\frac{m_{s}}{m_{b}}\sim\lambda^{2},\\
\nonumber &&\frac{m_{e}}{m_{\tau}}\sim
\lambda^{4},~~~~~~~~\frac{m_{\mu}}{m_{\tau}}\sim\lambda^{2}\\
\label{5}&&\frac{m_b}{m_t}\sim\lambda^{3}
\end{eqnarray}
Recent precision measurements have greatly improved the knowledge of
the CKM matrix, the experimental constraints on the CKM mixing
parameters are\cite{pdg}
\begin{equation}
\label{6}|V^{\rm{Exp}}_{\rm{CKM}}|\simeq\left(\begin{array}{ccc}
0.97377\pm0.00027&0.2257\pm0.0021&(4.31\pm0.30)\times10^{-3}\\
0.230\pm0.011&0.957\pm0.095&(41.6\pm0.6)\times10^{-3}\\
(7.4\pm0.8)\times10^{-3}&(40.6\pm2.7)\times10^{-3}&>0.78 \;at\;
95\%\;CL
\end{array}\right)
\end{equation}
The hierarchy in the quark mixing angles is clearly presented in the
Wolfenstein's parameterization of the CKM matrix\cite{pdg}.
Considering the scaling factor associated with the renormalization
group evolution of the CKM mixing angles from the electroweak scale
to the high scale, the magnitudes of the CKM matrix elements are
given in powers of $\lambda$ as follows
\begin{equation}
\label{7}
|V_{us}|\sim\lambda\,,~~~|V_{cb}|\sim\lambda^2\,,~~~|V_{td}|\sim\lambda^{3},~~~|V_{ub}|\sim\lambda^{4}
\end{equation}
Observations in the neutrino sector currently provide the strongest
indication for physics beyond the standard model. Including the new
data released by the MINOS and KamLAND collaborations, the global
fit of neutrino oscillation data at 2$\sigma$ indicates the
following values for the lepton mixing angles\cite{fit}
\begin{equation}
\label{8}0.28\leq\sin^2\theta_{12}\leq0.37,~~~0.38\leq\sin^2\theta_{23}\leq0.63,~~~\sin^2\theta_{13}\leq0.033
\end{equation}
and the best fit values are\cite{fit}
\begin{equation}
\label{9}\sin^2\theta_{12}=0.32,~~~\sin^2\theta_{23}=0.50,~~~\sin^2\theta_{13}=0.007
\end{equation}
The current data within 1$\sigma$ is well approximated by the
so-called TB mixing\cite{TBmix}
\begin{equation}
\label{10}U_{\rm{TB}}=\left(\begin{array}{ccc}
\sqrt{\frac{2}{3}}&\frac{1}{\sqrt{3}}&0\\
-\frac{1}{\sqrt{6}}&\frac{1}{\sqrt{3}}&-\frac{1}{\sqrt{2}}\\
-\frac{1}{\sqrt{6}}&\frac{1}{\sqrt{3}}&\frac{1}{\sqrt{2}}
\end{array}\right)
\end{equation}
which predicts $\sin^2\theta_{12,\rm{TB}}=\frac{1}{3}$,
$\sin^2\theta_{23,\rm{TB}}=\frac{1}{2}$ and
$\sin^2\theta_{13,\rm{TB}}=0$.

\section{brief review on the theory with U(2) flavor symmetry }

We shall briefly review the theory with U(2) flavor symmetry in the
following, which has been described in detail in the
literatures\cite{u2f}. The three generations of the matter fields
are assigned to transform as $\mathbf{2}\oplus\mathbf{1}$, the
sfermions of the first two generations are exactly degenerate in the
limit of unbroken U(2). In the low energy, this degeneracy is lifted
by the small symmetry breaking parameters which determine the light
fermion Yukawa couplings, therefore the flavor changing neutral
current(FCNC) and CP violating phenomena are sufficiently suppressed
so that the corresponding experimental bounds are not violated.
Three flavon fields $\phi^{a}$, $S^{ab}$ and $A^{ab}(a,b=1, 2)$ are
introduced, where $\phi$ is a U(2) doublet, $S$ and $A$ are
symmetric and antisymmetric tensors, and they are U(2) triplet and
singlet respectively. The hierarchies in the fermion masses and
mixing angles arise from the two step flavor symmetry breaking
\begin{equation}
\label{11}{\rm{U(2)}}\stackrel{\epsilon}{\rightarrow}{\rm{U(1)}}\stackrel{\epsilon'}{\rightarrow}nothing
\end{equation}
where both $\epsilon$ and $\epsilon'$ are small parameters with
$\epsilon>\epsilon'$. Both $\phi^{a}$ and $S^{ab}$ participate in
the first stage of symmetry breaking
${\rm{U(2)}}\stackrel{\epsilon}{\rightarrow}{\rm{U(1)}}$ with
$\langle\phi^1\rangle=0$, $\langle S^{11}\rangle=\langle
S^{12}\rangle=\langle S^{21}\rangle=0$, $\langle\phi^2\rangle={\cal
O}(\epsilon)$ and $\langle S^{22}\rangle={\cal O}(\epsilon)$. The
last stage of symmetry breaking is accomplished by $A^{ab}$ with
$\langle A^{12}\rangle=-\langle A^{21}\rangle={\cal O}(\epsilon')$.
The different mass hierarchies in the up sector and the down sector
can be understood by the combination of U(2) flavor symmetry and
grand unified symmetries\cite{u2f}, then the Yukawa matrices have
the following textures
\begin{eqnarray}
\nonumber Y_{U}&=&\left(\begin{array}{ccc}
0&\epsilon'\rho&0\\
-\epsilon'\rho&\epsilon\rho'&x_{u}\epsilon\\
0&y_{u}\epsilon&1
\end{array}\right)\zeta\\
\label{12}Y_{D,E}&=&\left(\begin{array}{ccc}
0&\epsilon'&0\\
-\epsilon'&(1,\pm3)\epsilon&(x_{d},x_e)\epsilon\\
0&(y_{d},y_{e})\epsilon&1
\end{array}\right)\varsigma
\end{eqnarray}
where $x_i,y_i={\cal O}(1)$ and $\varsigma\ll\zeta$. The model with
U(2) flavor symmetry successfully accounts for the quarks masses,
the charged lepton masses and the CKM mixing angles, and the
phenomenological constraints from FCNC and CP violation are
satisfied. It has been shown that the flavor models based on $T'$
symmetry could reproduce the Yukawa matrices in the U(2) flavor
theory\cite{qtp}, However, these models predicted the excluded small
mixing angle solution in the lepton sector. In the following we will
use triplet representation in the lepton sector to derive the TB
mixing naturally, singlet and doublet representations are exploited
in the quark sector, the Yukawa matrices in U(2) model are generated
at the leading order. Both the vacuum alignment and the next to
leading order corrections are discussed, which are crucial to the
flavor model building, however, these issues are omitted in
Ref.\cite{qtp}.

\section{the SUSY model with $T'\otimes Z_3\otimes Z_9$ flavor symmetry}

In our scheme, the symmetry group is $SU(3)_c\otimes SU(2)_L\otimes
U(1)_Y\otimes G_F$, where $G_F$ is the global flavor symmetry group
$G_F=T'\otimes Z_3\otimes Z_9$.  The $Z_3$ symmetry is to guarantee
the correct misalignment in flavor space between the neutrino masses
and the charged lepton masses as in
Ref.\cite{Altarelli:2005yx,Feruglio:2007uu}, and $Z_9$ is crucial to
obtain the realistic hierarchies in the fermion masses and mixing
angles. In addition to the minimal supersymmtric standard model
(MSSM) matter fields, we need to introduce the fields which are
responsible for the flavor symmetry breaking, we refer to these
fields as flavons which are gauge singlets. Both the MSSM fields and
the flavon fields and their transformation properties under
$T'\otimes Z_3\otimes Z_9$ are shown in Table \ref{table1}, where
$\alpha$ and $\beta$ are respectively the generators of $Z_3$ and
$Z_9$ with $\alpha=\exp[i2\pi/3]$ and $\beta=\exp[i2\pi/9]$. Note
that although the flavons $\theta'$ and $\chi$ are not involved in
the leading order Yukawa superpotential, they play an important role
in the vacuum alignment mechanism.

\begin{table}[hptb]
\begin{center}
\begin{tabular}{|c|c|c|c|c|c|c|c|c|c|c|c||c|c|c|c|c|c|c|c|c|}\hline\hline
Fields& $~\ell~$  & $~e^{c}~$ & $~\mu^{c}~$ &  $~\tau^{c}~$   &
$Q_L$ & $U^{c}$ & $D^{c}$& $~Q_3~$ & $~t^{c}~$ & $~b^{c}~$ &
$H_{u,d}$& $\varphi_{T}$  & $\varphi_{S}$ & $~\xi,\tilde{\xi}~$ &
~$\phi$~ & ~$\theta''$~ & ~$\theta'$~ &~$\Delta$~ & ~$\bar{\Delta}$~
& ~$\chi$~\\\hline

$T'$& 3  & $\mathbf{1}$& $\mathbf{1}''$ & $\mathbf{1}'$ &
$\mathbf{2}'$ & $\mathbf{2}$& $\mathbf{2}$ & $\mathbf{1}''$ &
$\mathbf{1}'$ & $\mathbf{1}'$& $\mathbf{1}$& $\mathbf{3}$
&$\mathbf{3}$& $\mathbf{1}$ & $\mathbf{2}'$ & $\mathbf{1}''$ &
$\mathbf{1}'$ & $\mathbf{1}$&$\mathbf{1}$ &$\mathbf{1}$\\\hline

$Z_{3}$& $\alpha$ & $\alpha^2$  & $\alpha^2$ & $\alpha^2$ & $\alpha$
&$\alpha^2$  & $\alpha^2$ &$\alpha$  & $\alpha^2$  &  $\alpha^2$
&$\mathbf{1}$& $\mathbf{1}$ & $\alpha$ & $\alpha$ &  $\mathbf{1}$  &
$\mathbf{1}$ & $\mathbf{1}$ & $\mathbf{1}$ & $\mathbf{1}$ &
$\mathbf{1}$
\\\hline

$Z_{9}$&$\mathbf{1}$ &$\mathbf{1}$ & $\beta^6$ & $\beta^8$ &
$\beta^3$  & $\beta^3$& $\beta$& $\mathbf{1}$ & $\mathbf{1}$ &
$\beta^7$ &$\mathbf{1}$& $\beta$ & $\mathbf{1}$ & $\mathbf{1}$ &
$\beta^6$ & $\beta$ &$\beta$ & $\beta^2$& $\beta^4$
&$\beta$\\\hline\hline
\end{tabular}
\end{center}
\caption{\label{table1}The transformation rules of the MSSM fields
and the flavon fields under the flavor symmetry $T'\otimes
Z_3\otimes Z_9$. We denote $Q_L=(Q_1,Q_2)^T$, where
$Q_1=(u_L,d_L)^T$ and $Q_2=(c_L,s_L)^T$ are the electroweak
$SU(2)_L$ doublets of the first two generations. $U^c=(u^c,c^c)^T$
and $D^c=(d^c,s^c)^T$, $Q_L$, $U^c$ and $D^c$ are $T'$ doublets.
$Q_3=(t_L,b_L)^T$ is the electroweak $SU(2)_L$ doublet of the third
generation, $Q_3$, $t^c$ and $b^c$ are $T'$ singlets . The up type
and down type Higgs transform as a singlet under the flavor group. }
\end{table}

As we shall demonstrate in section V, at the leading order, the
scalar components of the flavon supermultiplets $\varphi_{T}$,
$\varphi_{S}$ etc. develop vacuum expectation values(VEV) along the
following diractions
\begin{eqnarray}
\nonumber &&\langle \varphi_{T}\rangle=(v_{T},0,0),~~~\langle
\varphi_{S}\rangle=(v_S,v_S,v_S),~~~\langle\phi\rangle=(v_1,0),\\
\nonumber&&\langle\xi\rangle=u_{\xi},~~~\langle\tilde{\xi}\rangle=0,~~~\langle\theta'\rangle=u'_{\theta},~~~\langle\theta''\rangle=u''_{\theta},\\
\label{13}&&\langle\Delta\rangle=u_{\Delta},~~~\langle\bar{\Delta}\rangle=\bar{u}_{\Delta},~~~\langle\chi\rangle=u_{\chi}
\end{eqnarray}
The electroweak symmetry is broken by the up and down type Higgs
with $\langle H_{u,d}\rangle=v_{u,d}$. As we shall see in the
following, in order to obtain the realistic pattern of charged
fermion masses and mixing angles, these VEVs should be of the orders
\begin{equation}
\label{14}|\frac{v_{T}}{\Lambda}|\approx|\frac{v_{S}}{\Lambda}|\approx|\frac{v_1}{\Lambda}|\sim\lambda^2,~~~~|\frac{u'_{\theta}}{\Lambda}|\approx|\frac{u''_{\theta}}{\Lambda}|\approx|\frac{u_{\Delta}}{\Lambda}|\approx|\frac{\bar{u}_{\Delta}}{\Lambda}|\sim\lambda^3
\end{equation}
where $\Lambda$ is the cut off scale of the theory, these relations
imply that the VEVs of the $T'$ triplets and doublet are required to
be of order $\lambda^2\Lambda$, while the VEVs of the $T'$ singlets
$\theta$, $\theta'$, $\Delta$ and $\bar{\Delta}$ are of order
$\lambda^3\Lambda$. Naturally $u_{\xi}$ and $u_{\chi}$ should be of
the order $\lambda^2\Lambda\sim\lambda^3\Lambda$ as well. The VEVs
of required orders in Eq.(\ref{14}) can be achieved in a finite
portion of the parameter space, which will be illustrated in the
discussion of the vacuum alignment.

\subsection{The lepton sector}
The Yukawa interactions in the lepton sector are controlled by the
superpotential
\begin{equation}
\label{15}w_{\ell}=w_{e}+w_{\nu}
\end{equation}
where we have separated the contribution to the neutrino masses and
the charged lepton masses, both $w_{e}$ and $w_{\nu}$ are invariant
under the gauge group of the standard model and the flavor symmetry
$T'\otimes Z_3\otimes Z_9$. The leading order terms of the Yukawa
superpotential $w_{e}$ are
\begin{eqnarray}
\nonumber&&w_{e}=y_ee^{c}(\ell\varphi_T)\bar{\Delta}^2H_{d}/\Lambda^3+h_{e1}e^{c}(\ell\varphi_S)(\varphi_S\varphi_S)H_{d}/\Lambda^3+h_{e2}e^{c}(\ell\varphi_s)'(\varphi_S\varphi_S)''H_{d}/\Lambda^3\\
\nonumber&&+h_{e3}e^{c}(\ell\varphi_S)''(\varphi_S\varphi_S)'H_{d}/\Lambda^3+h_{e4}e^{c}(\ell\varphi_S)\xi^2H_d/\Lambda^3+y_{\mu1}\mu^{c}(\ell\phi\phi)'H_{d}/\Lambda^2+y_{\mu2}\mu^{c}(\ell\varphi_T)'\Delta
H_{d}/\Lambda^2\\
\nonumber&&+h_{\mu1}\mu^{c}(\ell\varphi_T)'(\varphi_T\varphi_T)H_{d}/\Lambda^3+h_{\mu2}\mu^{c}((\ell\varphi_T)_{\mathbf{3}_{S}}(\varphi_T\varphi_T)_{\mathbf{3}_S})'H_{d}/\Lambda^3+h_{\mu3}\mu^{c}((\ell\varphi_T)_{\mathbf{3}_{A}}(\varphi_T\varphi_T)_{\mathbf{3}_S})'H_{d}/\Lambda^3\\
\nonumber&&+h_{\mu4}\mu^{c}(\ell\varphi_T\varphi_T)'\chi H_{d}/\Lambda^3+h_{\mu5}\mu^{c}(\ell\varphi_T\varphi_T)\theta'H_{d}/\Lambda^3+h_{\mu6}\mu^{c}(\ell\varphi_T\varphi_T)''\theta''H_{d}/\Lambda^3+h_{\mu7}\mu^{c}(\ell\varphi_T)'\chi^2H_{d}/\Lambda^3\\
\nonumber&&+h_{\mu8}\mu^{c}(\ell\varphi_T)'\theta'\theta''H_{d}/\Lambda^3+h_{\mu9}\mu^{c}(\ell\varphi_T)\chi\theta'H_{d}/\Lambda^3+h_{\mu10}\mu^{c}(\ell\varphi_T)\theta''\theta''H_{d}/\Lambda^3+h_{\mu11}\mu^{c}(\ell\varphi_T)''\chi\theta''H_{d}/\Lambda^3\\
\label{16}&&+h_{\mu12}\mu^{c}(\ell\varphi_T)''\theta'\theta'H_{d}/\Lambda^3+y_{\tau}\tau^{c}(\ell\varphi_T)''H_{d}/\Lambda+...
\end{eqnarray}
where dots stand for additional operators of order $1/\Lambda^3$,
whose contributions to the charged lepton masses vanish at the
leading order. The coefficients $y_{e}$, $h_{ei}(i=1,2,3,4)$,
$y_{\mu1}$, $y_{\mu2}$, $h_{\mu i}(i=1-12)$ and $y_{\tau}$ are
naturally ${\cal O}(1)$ coupling constants. After the electroweak
symmetry breaking and the flavor symmetry breaking, the charged
lepton mass terms from $w_{e}$ are
\begin{eqnarray}
\nonumber&&w_{e}=y_{e}\frac{\bar{u}^2_{\Delta}v_{T}}{\Lambda^3}v_de^{c}e+[3(h_{e1}+h_{e2}+h_{e3})\frac{v^3_S}{\Lambda^3}+h_{e4}\frac{u^2_{\xi}v_S}{\Lambda^3}]v_de^{c}(e+\mu+\tau)\\
\nonumber&&+(iy_{\mu1}\frac{v^2_1}{\Lambda^2}+y_{\mu2}\frac{u_{\Delta}v_T}{\Lambda^2})v_d\mu^c\mu+(h_{\mu5}\frac{2u'_{\theta}v^2_T}{3\Lambda^3}+h_{\mu9}\frac{u_{\chi}u'_{\theta}v_T}{\Lambda^3}+h_{\mu10}\frac{u''^{2}_{\theta}v_T}{\Lambda^3})v_d\mu^{c}e\\
\nonumber&&+[(h_{\mu1}-\frac{2}{9}h_{\mu2}-\frac{1}{3}h_{\mu3})\frac{v^3_T}{\Lambda^3}+h_{\mu4}\frac{2u_{\chi}v^2_T}{3\Lambda^3}+h_{\mu7}\frac{u^2_{\chi}v_T}{\Lambda^3}+h_{\mu8}\frac{u'_{\theta}u''_{\theta}v_T}{\Lambda^3}]v_d\mu^{c}\mu\\
\nonumber&&+(h_{\mu6}\frac{2u''_{\theta}v^2_T}{3\Lambda^3}+h_{\mu11}\frac{u_{\chi}u''_{\theta}v_T}{\Lambda^3}+h_{\mu12}\frac{u'^{2}_{\theta}v_T}{\Lambda^3})v_d\mu^{c}\tau+y_{\tau}\frac{v_{T}}{\Lambda}v_{d}\tau^{c}\tau\\
\nonumber&&\equiv
(y_{e}\frac{\bar{u}^2_{\Delta}v_{T}}{\Lambda^3}+y'_e\frac{v^3_S}{\Lambda^3})v_de^{c}e+y'_e\frac{v^3_S}{\Lambda^3}v_de^{c}\mu+y'_e\frac{v^3_S}{\Lambda^3}v_de^{c}\tau+y_{\mu
e}\frac{u'_{\theta}v^2_T}{\Lambda^3}v_d\mu^ce+y_{\mu}\frac{v^2_1}{\Lambda^2}v_d\mu^c\mu\\
\label{17}&&+y_{\mu\tau}\frac{u''_{\theta}v^2_T}{\Lambda^3}v_d\mu^{c}\tau+y_{\tau}\frac{v_T}{\Lambda}v_d\tau^{c}\tau
\end{eqnarray}
where $y'_{e}=3(h_{e1}+h_{e2}+h_{e3})+h_{e4}\frac{u^2_{\xi}}{v^2_S}$
, $y_{\mu}\approx iy_{\mu1}+y_{\mu2}\frac{u_{\Delta}v_T}{v^2_1}$,
$y_{\mu
e}=\frac{2}{3}h_{\mu5}+h_{\mu9}\frac{u_{\chi}}{v_T}+h_{\mu10}\frac{u''^2_{\theta}}{u'_{\theta}v_T}$
and
$y_{\mu\tau}=\frac{2}{3}h_{\mu6}+h_{\mu11}\frac{u_{\chi}}{v_T}+h_{\mu12}\frac{u'^2_{\theta}}{u''_{\theta}v_T}$.
Therefore at the leading order, the charged lepton mass matrix is
given by
\begin{equation}
\label{18}M^{e}=\left(\begin{array}{ccc}
y_e\frac{\bar{u}^{2}_{\Delta}v_{T}}{\Lambda^3}+y'_{e}\frac{v^3_S}{\Lambda^3}&y'_{e}\frac{v^3_S}{\Lambda^3}&y'_{e}\frac{v^3_S}{\Lambda^3}\\
y_{\mu e}\frac{u'_{\theta}v^2_T}{\Lambda^3}&y_{\mu}\frac{v^2_1}{\Lambda^2}&y_{\mu\tau}\frac{u''_{\theta}v^2_T}{\Lambda^3}\\
0&0&y_{\tau}\frac{v_{T}}{\Lambda}
\end{array}\right)v_d
\end{equation}
Note that the charged lepton mass matrix is no longer diagonal at
the leading order in contrast with Ref.
\cite{Altarelli:2005yx,Feruglio:2007uu}. Since the charged lepton
masses receive contribution from the VEV of $\varphi_S$, $T'$ is
completely broken already at the leading order. Whereas $T'$ is
broken down to $Z_3$ at the leading order, then it is broken to
nothing by the higher dimensional operators in
Ref.\cite{Altarelli:2005yx,Feruglio:2007uu}. The mass matrix $M^{e}$
is diagonalized by a biunitary transformation
$V^{e\dagger}_{R}M^{e}V^{e}_{L}=diag(m_e,m_{\mu},m_{\tau})$,
therefore
$V^{e\dagger}_{L}M^{e\dagger}M^{e}V^{e}_{L}=diag(m^2_e,m^2_{\mu},m^2_{\tau})$.
The matrix $V^{e}_L$ approximately is
\begin{equation}
\label{19}V^{e}_L\approx\left(\begin{array}{ccc}
1& s^{e}_{12}&0\\
-s^{e*}_{12}&1&0\\
0&0&1
\end{array}\right)
\end{equation}
where $s^{e}_{12}=(\frac{y_{\mu
e}}{y_{\mu}}\frac{u'_{\theta}v^2_T}{v^2_1\Lambda})^*+\frac{|y'_{e}|^2}{|y_{\mu}|^2}\frac{|v_S|^6}{|v_{1}|^4\Lambda^2}$,
and the charged lepton masses are approximately given by
\begin{eqnarray}
\nonumber
m_{e}&\approx&\Big|(y_e\frac{\bar{u}^2_{\Delta}v_T}{\Lambda^3}+y'_e\frac{v^3_S}{\Lambda^3})v_d\Big|\\
\nonumber
m_{\mu}&\approx&\Big|y_{\mu}\frac{v^2_1}{\Lambda^2}v_d\Big|\\
\label{20}m_{\tau}&\approx&\Big|y_{\tau}\frac{v_T}{\Lambda}v_d\Big|
\end{eqnarray}
Therefore the mass ratios are estimated
\begin{equation}
\label{21}
\frac{m_e}{m_{\tau}}\approx\Big|\frac{y_{e}}{y_{\tau}}\frac{\bar{u}_{\Delta}}{\Lambda^2}+\frac{y'_{e}}{y_{\tau}}\frac{v^3_S}{v_{T}\Lambda^2}\Big|\approx\Big|\frac{y'_{e}}{y_{\tau}}\frac{v^3_S}{v_{T}\Lambda^2}\Big|,~~~~~\frac{m_{\mu}}{m_{\tau}}\approx\Big|\frac{y_{\mu}}{y_{\tau}}\frac{v^2_1}{v_T\Lambda}\Big|
\end{equation}
From Eq.(\ref{14}) and Eq.(\ref{21}), we see that the realistic
hierarchies among the charged lepton masses
$m_{\tau}:m_{\mu}:m_{e}\approx1:\lambda^2:\lambda^4$ are produced
naturally. For the neutrino sector, we have
\begin{equation}
\label{23}w_{\nu}=(y_{\xi}\xi+\tilde{y}_{\xi}\tilde{\xi})(\ell\ell)H_{u}H_{u}/\Lambda^2+y_{S}(\varphi_S\ell\ell)H_{u}H_{u}/\Lambda^2+...
\end{equation}
after the electroweak and flavor symmetry breaking, $w_{\nu}$ gives
rise to the following mass terms for the neutrinos
\begin{equation}
\label{24}w_{\nu}=y_{\xi}\frac{u_{\xi}}{\Lambda}\frac{v^2_u}{\Lambda}(\nu^2_e+2\nu_{\mu}\nu_{\tau})+\frac{2}{3}y_{S}\frac{v_S}{\Lambda}\frac{v^2_u}{\Lambda}(\nu^2_e+\nu^2_{\mu}+\nu^2_{\tau}-\nu_e\nu_{\mu}-\nu_{e}\nu_{\tau}-\nu_{\mu}\nu_{\tau})+...
\end{equation}
Therefore at the leading order the neutrino mass matrix is
\begin{equation}
\label{25}M^{\nu}=\left(\begin{array}{ccc}
2y_{\xi}\frac{u_{\xi}}{\Lambda}+\frac{4}{3}y_S\frac{v_S}{\Lambda}&-\frac{2}{3}y_S\frac{v_S}{\Lambda} &-\frac{2}{3}y_S\frac{v_S}{\Lambda}\\
-\frac{2}{3}y_S\frac{v_S}{\Lambda}&\frac{4}{3}y_S\frac{v_S}{\Lambda}&2y_{\xi}\frac{u_{\xi}}{\Lambda}-\frac{2}{3}y_S\frac{v_S}{\Lambda}\\
-\frac{2}{3}y_S\frac{v_S}{\Lambda}&2y_{\xi}\frac{u_{\xi}}{\Lambda}-\frac{2}{3}y_S\frac{v_S}{\Lambda}&\frac{4}{3}y_S\frac{v_S}{\Lambda}
\end{array}\right)\frac{v^2_u}{\Lambda}
\end{equation}
$M^{\nu}$ is diagonalized by a unitary transformation $V^{\nu}_{L}$
\begin{equation}
\label{26}V^{\nu
T}_LM^{\nu}V^{\nu}_L=diag(2y_{\xi}\frac{u_{\xi}}{\Lambda}+2y_{S}\frac{v_S}{\Lambda},2y_{\xi}\frac{u_{\xi}}{\Lambda},-2y_{\xi}\frac{u_{\xi}}{\Lambda}+2y_{S}\frac{v_S}{\Lambda})\frac{v^2_u}{\Lambda}
\end{equation}
Where the diagonalization matrix $V^{\nu}_{L}$ is the tri-bimaximal
mixing matrix $V^{\nu}_{L}=U_{TB}$, therefore the
Maki-Nakagawa-Sakata-Pontecorvo(MNSP) mixing matrix, at this order,
is
\begin{equation}
\label{27}V_{\rm{MNSP}}=V^{e\;\dagger}_LV^{\nu}_L\approx\left(\begin{array}{ccc}
\sqrt{\frac{2}{3}}+\frac{1}{\sqrt{6}}s^{e}_{12}&\frac{1}{\sqrt{3}}-\frac{1}{\sqrt{3}}s^{e}_{12}&\frac{1}{\sqrt{2}}s^{e}_{12}\\
-\frac{1}{\sqrt{6}}+\sqrt{\frac{2}{3}}s^{e*}_{12}&\frac{1}{\sqrt{3}}+\frac{1}{\sqrt{3}}s^{e*}_{12}&-\frac{1}{\sqrt{2}}\\
-\frac{1}{\sqrt{6}}&\frac{1}{\sqrt{3}}&\frac{1}{\sqrt{2}}
\end{array}\right)
\end{equation}
We see that the MNSP matrix deviates from the TB mixing pattern due
to the corrections from the charged lepton sector, in particular,
$(V_{\rm{MNSP}})_{e3}$ is no longer identically zero
\begin{eqnarray}
\nonumber|(V_{\rm
MNSP})_{e3}|&\approx&\frac{1}{\sqrt{2}}|s^e_{12}|=\frac{1}{\sqrt{2}}\Big|\big(\frac{y_{\mu
e}}{y_{\mu}}\frac{u'_{\theta}v^2_T}{v^2_1\Lambda}\big)^*+\frac{|y'_{e}|^2}{|y_{\mu}|^2}\frac{|v_S|^6}{|v_{1}|^4\Lambda^2}\Big|\\
\nonumber\tan^2\theta_{23}&\approx&1\\
\label{28}\tan^2\theta_{12}&\approx&\frac{1}{2}-\frac{3}{4}\Big[\frac{y_{\mu
e}}{y_{\mu}}\frac{u'_{\theta}v^2_T}{v^2_1\Lambda}+\big(\frac{y_{\mu
e}}{y_{\mu}}\frac{u'_{\theta}v^2_T}{v^2_1\Lambda}\big)^*+2\frac{|y'_{e}|^2}{|y_{\mu}|^2}\frac{|v_S|^6}{|v_{1}|^4\Lambda^2}\Big]
\end{eqnarray}
From Eq.(\ref{14}), we learn that $s^{e}_{12}$ is of order
$\lambda^3$, therefore at leading order the MNSP matrix is very
close to the TB mixing matrix, and the corrections from the charged
lepton sector are very small.

\subsection{The quark sector}
The Yukawa interactions in the quark sector are
\begin{equation}
\label{29}w_{q}=w_{u}+w_{d}
\end{equation}
For the up quark sector, we have
\begin{eqnarray}
\nonumber w_{u}&=&y_{u1}(\varphi_TQ_LU^{c})\Delta
H_{u}/\Lambda^2+y_{u2}((Q_LU^{c})_{\mathbf{3}}(\phi\phi)_{\mathbf{3}})H_{u}/\Lambda^2+y_{u3}(Q_LU^{c})'\theta''\Delta
H_{u}/\Lambda^2\\
\label{30}&&+y_{u4}(Q_L\phi)''t^{c}H_{u}/\Lambda+y_{u5}Q_3(U^c\phi)'H_{u}/\Lambda+y_tQ_3t^{c}H_{u}+...
\end{eqnarray}
In the down quark sector, we obtain
\begin{eqnarray}
\nonumber
w_d&=&y_{d1}(\varphi_TQ_LD^c)\bar{\Delta}H_{d}/\Lambda^2+y_{d2}(Q_LD^{c})'\theta''\bar{\Delta}H_{d}/\Lambda^2+y_{d3}(Q_L\phi)''b^{c}\Delta
H_{d}/\Lambda^2\\
\nonumber&&+y_{d4}Q_3(D^{c}\phi)'\Delta
H_{d}/\Lambda^2+y_{b1}Q_3b^{c}\Delta
H_{d}/\Lambda+y_{b2}Q_3b^c(\varphi_T\varphi_T)h_d/\Lambda^2+y_{b3}Q_3b^c\chi^2h_d/\Lambda^2\\
\label{31}&&+y_{b4}Q_3b^c\theta'\theta''/\Lambda^2...
\end{eqnarray}
After electroweak and flavor symmetry breaking, we have the quark
mass terms
\begin{eqnarray}
\nonumber w_{q}&=&y_{u1}\frac{u_{\Delta}v_{T}}{\Lambda^2}v_{u}cc^{c}+iy_{u2}\frac{v^2_1}{\Lambda^2}v_{u}cc^{c}+y_{u3}\frac{u''_{\theta}u_{\Delta}}{\Lambda^2}v_u(uc^c-cu^c)+y_{u4}\frac{v_1}{\Lambda}v_uct^{c}+y_{u5}\frac{v_1}{\Lambda}v_utc^{c}\\
\nonumber&&+y_tv_utt^c+y_{d1}\frac{\bar{u}_{\Delta}v_T}{\Lambda^2}v_dss^{c}+y_{d2}\frac{u''_{\theta}\bar{u}_{\Delta}}{\Lambda^2}v_d(ds^{c}-sd^{c})+y_{d3}\frac{u_{\Delta}v_1}{\Lambda^2}v_dsb^{c}+y_{d4}\frac{u_{\Delta}v_1}{\Lambda^2}v_dbs^{c}\\
\label{32}&&+y_{b}\frac{u_{\Delta}}{\Lambda}v_dbb^{c}
\end{eqnarray}
where
$y_b=y_{b1}+y_{b2}\frac{v^2_T}{u_{\Delta}\Lambda}+y_{b3}\frac{u^2_{\chi}}{u_{\Delta}\Lambda}+y_{b4}\frac{u'_{\theta}u''_{\theta}}{u_{\Delta}\Lambda}$,
and the resulting quark mass matrices are
\begin{eqnarray}
\nonumber M^{u}&=&\left(\begin{array}{ccc}
0&-y_{u3}\frac{u''_{\theta}u_{\Delta}}{\Lambda^2}&0\\
y_{u3}\frac{u''_{\theta}u_{\Delta}}{\Lambda^2}&y_{u1}\frac{u_{\Delta}v_{T}}{\Lambda^2}+iy_{u2}\frac{v^2_1}{\Lambda^2}&y_{u5}\frac{v_1}{\Lambda}\\
0&y_{u4}\frac{v_1}{\Lambda}&y_t
\end{array}\right)v_u\\
\label{33}M^{d}&=&\left(\begin{array}{ccc}
0&-y_{d2}\frac{u''_{\theta}\bar{u}_{\Delta}}{\Lambda^2}&0\\
y_{d2}\frac{u''_{\theta}\bar{u}_{\Delta}}{\Lambda^2}&y_{d1}\frac{\bar{u}_{\Delta}v_{T}}{\Lambda^2}&y_{d4}\frac{u_{\Delta}v_1}{\Lambda^2}\\
0&y_{d3}\frac{u_{\Delta}v_1}{\Lambda^2}&y_b\frac{u_{\Delta}}{\Lambda}
\end{array}\right)v_d
\end{eqnarray}
We see that both $M^{u}$ and $M^{d}$ have the same textures as those
in the U(2) flavor model\cite{u2f}. From the Appendix A, we see that
under the $T'$ generator $T$, the quark fields transform as
$Q_1\stackrel{T}{\longrightarrow}Q_1$,
$Q_2(Q_3,u^c,d^c)\stackrel{T}{\longrightarrow}\omega^2
Q_2(Q_3,u^c,d^c)$ and
$c^c(t^c,s^c,b^c)\stackrel{T}{\longrightarrow}\omega
c^c(t^c,s^c,b^c)$. Consequently, if the vacuum expectation value of
$\theta''$ vanishes, the above mass matrices are the most general
ones invariant under the subgroup $Z^{''}_3$ generated by the
generator $T$. In this work, $u''_{\theta}$ further breaks $Z_3$ to
nothing. Diagonalizing the quark mass matrices in Eq.(\ref{33})
using the standard perturbation
technique\cite{Hall:1993ni,Leurer:1992wg}, we obtain the quark
masses as follows
\begin{eqnarray}
\nonumber
m_{u}&\approx&\Big|\frac{y^{2}_{u3}y_tu''^{2}_{\theta}u^2_{\Delta}}{(iy_{u2}y_t-y_{u4}y_{u5})v^2_1\Lambda^2}v_u\Big|\\
\nonumber
m_{c}&\approx&\Big|(iy_{u2}-\frac{y_{u4}y_{u5}}{y_t})\frac{v^2_1}{\Lambda^2}v_u\Big|\\
\nonumber m_t&\approx&|y_tv_u|\\
\nonumber
m_{d}&\approx&\Big|\frac{y^2_{d2}u''^{2}_{\theta}\bar{u}_{\Delta}}{y_{d1}v_{T}\Lambda^2}v_d\Big|\\
\nonumber m_s&\approx&\Big|y_{d1}\frac{\bar{u}_{\Delta}v_{T}}{\Lambda^2}v_d\Big|\\
\label{34}m_{b}&\approx&\Big|y_b\frac{u_{\Delta}}{\Lambda}v_d\Big|
\end{eqnarray}
and the CKM matrix elements are estimated as
\begin{eqnarray}
\nonumber&&V_{ud}\approx V_{cs}\approx V_{tb}\approx1\\
\nonumber&&V^{*}_{us}\approx
-V_{cd}\approx\frac{y_{d2}}{y_{d1}}\frac{u''_{\theta}}{v_{T}}-\frac{y_{u3}y_tu''_{\theta}u_{\Delta}}{(iy_{u2}y_t-y_{u4}y_{u5})v^2_1}\\
\nonumber&&V^{*}_{cb}\approx-V_{ts}\approx(\frac{y_{d3}}{y_b}-\frac{y_{u4}}{y_t})\frac{v_1}{\Lambda}\\
\nonumber&&V^{*}_{ub}\approx
-\frac{y_{u3}y_t}{iy_{u2}y_t-y_{u4}y_{u5}}(\frac{y_{d3}}{y_b}-\frac{y_{u4}}{y_{t}})\frac{u''_{\theta}u_{\Delta}}{v_1\Lambda}+\frac{y_{d2}y^{*}_{d4}}{|y_b|^2}\frac{u''_{\theta}\bar{u}_{\Delta}v^{*}_1}{u_{\Delta}\Lambda^2}\\
\label{35}&&V_{td}\approx\frac{y_{d2}}{y_{d1}}(\frac{y_{d3}}{y_{b}}-\frac{y_{u4}}{y_t})\frac{u''_{\theta}v_1}{v_{T}\Lambda}-\frac{y_{d2}y^{*}_{d4}}{|y_b|^2}\frac{u''_{\theta}\bar{u}_{\Delta}v^{*}_1}{u_{\Delta}\Lambda^2}
\end{eqnarray}
From Eq.(\ref{14}) and Eq.(\ref{34}), we see that the correct quark
mass hierarchies are reproduced
$m_t:m_c:m_u\sim1:\lambda^4:\lambda^8$,
$m_b:m_s:m_d\sim1:\lambda^2:\lambda^4$ and
$m_t:m_{b}\sim1:\lambda^3$. Moreover, Eq.(\ref{20}) and
Eq.(\ref{34}) imply that the tau lepton and bottom quark masses are
respectively of the order $\lambda^2$ and $\lambda^3$. Since
$b-\tau$ unification $m_b\simeq m_{\tau}$ is usually predicted in
many unification models, we expect to achieve $b-\tau$ unification
in GUT model with $T'$ flavor symmetry as well, without changing
drastically the successful predictions for flavor mixings and
fermion mass hierarchies presented here\cite{ding}. In our model
$\tan\beta\equiv v_u/v_{d}$ is of order one, the hierarchy between
the top quark and bottom quark masses is due to the flavor symmetry
breaking pattern. However, in Ref.\cite{Feruglio:2007uu} the large
mass difference between the top and bottom quark is due to large
$\tan\beta$, consequently there are large radiative corrections to
the quark masses and the CKM matrix elements, which may
significantly alter the low energy predictions of quark masses and
CKM matrix. From Eq.(\ref{14}) and Eq.(\ref{35}), we learn that the
correct hierarchy of the CKM matrix elements in Eq.(\ref{7}) is
generated as well. Two interesting relations between the quark
masses and mixing angles are predicted
\begin{equation}
\label{36}
\Big|\frac{V_{td}}{V_{ts}}\Big|\approx\sqrt{\frac{m_d}{m_s}}\,,~~~~\Big|\frac{V_{ub}}{V_{cb}}\Big|\approx\sqrt{\frac{m_u}{m_c}}
\end{equation}
The above relations are also predicted in U(2) flavor theory. The
first relation is satisfied within the large theoretical errors of
both sides, and the second relation is not so well fulfilled as the
first one. Both relations will be corrected by the next to leading
order operators.

\section{vacuum alignment}
In section IV we have demonstrated that the realistic pattern of
fermion masses and flavor mixing are generated, if $T'$ is broken
along the directions shown in Eq.(\ref{13}), in the following we
will illustrate that the VEVs in Eq.(\ref{13}) is really a local
minimum of the scalar potential of the model in a finite portion of
the parameter space. Using the technique in
Ref.\cite{Altarelli:2005yx,Bazzocchi:2007na,Feruglio:2007uu}, a
global continuous $U(1)_R$ symmetry is exploited to simplify the
vacuum alignment problem, and this symmetry is broken to the
discrete R-parity once we include the gaugino masses in the model.
The Yukawa superpotentials $w_{\ell}$ and $w_{q}$ in Eq.(\ref{15})
and Eq.(\ref{29}) are invariant under the $U(1)_R$ symmetry, if +1
R-charge is assigned to the matter fields (i.e. the lepton and quark
superfields), and 0 R-charge to the Higgs and flavon
supermultiplets. Since the superpotential must have +2 R-charge, we
should introduce some driving fields which carry +2 R-charge in
order to avoid the spontaneous breaking of the $U(1)_R$ symmetry,
consequently the driving fields enter linearly into the terms of the
superpotential. The driving fields and and their transformation
properties under $T'\otimes Z_3\otimes Z_9$ are shown in Table
\ref{table2}.

\begin{table}[hptb]
\begin{center}
\begin{tabular}{|c|c|c|c|c|c|c|c|c|}\hline\hline
Fields& $\varphi^{R}_{T}$  & $\varphi^{R}_{S}$ & $~\xi^{R}~$ &
~$\phi^{R}$~ & ~$\theta''^{R}$~ & ~$\Delta^R$~ &~$\bar{\Delta}^R$~ &
~$\chi^{R}$~ \\\hline

$T'$& $\mathbf{3}$  & $\mathbf{3}$& $\mathbf{1}$ & $\mathbf{2}''$ &
$\mathbf{1}''$ & $\mathbf{1}$ & $\mathbf{1}$ & $\mathbf{1}$
\\\hline

$Z_{3}$& $\mathbf{1}$  & $\alpha$ & $\alpha$ & $\mathbf{1}$   &
$\mathbf{1}$  & $\mathbf{1}$ & $\mathbf{1}$ & $\mathbf{1}$
\\\hline

$Z_{9}$& $\beta^6$ & $\mathbf{1}$ & $\mathbf{1}$ & $\beta^2$   &
$\beta^7$ & $\beta^7$ & $\beta^5$ & $\beta^7$
\\\hline\hline
\end{tabular}
\end{center}
\caption{\label{table2}The driving fields and their transformation
rules under $T'\otimes Z_3\otimes Z_9$}
\end{table}

At the leading order, the superpotential depending on the driving
fields, which is invariant under all the symmetry of the model, is
given by
\begin{eqnarray}
\nonumber&&w_{v}=g_1(\varphi^{R}_T\phi\phi)+g_2(\varphi^{R}_T\varphi_T)\Delta+g_3(\phi^R\phi)\chi+g_4(\varphi_T\phi^{R}\phi)+g_5\chi^R\chi^2+g_6\chi^R\theta'\theta''\\
\nonumber&&+g_7\chi^R(\varphi_T\varphi_T)+g_8\theta''^R\theta''^2+g_9\theta''^R\theta'\chi+g_{10}\theta''^R(\varphi_T\varphi_T)'+M_{\Delta}\Delta^R\Delta+g_{11}\Delta^{R}\chi^2\\
\nonumber&&+g_{12}\Delta^{R}\theta'\theta''+g_{13}\Delta^R(\varphi_T\varphi_T)+\bar{M}_{\Delta}\bar{\Delta}^R\bar{\Delta}+g_{14}\bar{\Delta}^{R}\Delta^2+g_{15}(\varphi^{R}_{S}\varphi_S\varphi_S)+g_{16}(\varphi^R_S\varphi_S)\tilde{\xi}\\
\label{37}&&+g_{17}\xi^R(\varphi_S\varphi_S)+g_{18}\xi^R\xi^2+g_{19}\xi^R\xi\tilde{\xi}+g_{20}\xi^R\tilde{\xi}^2
\end{eqnarray}
Since there is no distinction between $\xi$ and $\tilde{\xi}$, we
define $\tilde{\xi}$ as the field that couples to
$(\varphi^R_S\varphi_S)$ in the superpotential $w_{v}$ as in
Ref.\cite{Altarelli:2005yx,Feruglio:2007uu}, and $\tilde{\xi}$ is
necessary to achieve the correct vacuum alignment. Similarly the
quantum numbers of $\Delta^R$ and $\chi^R$  are exactly identical,
we define $\Delta^R$ as the one which couples to $\Delta$.

From the superpotential $w_{v}$ in Eq.(\ref{37}), we can derive the
scalar potential of this model
\begin{equation}
\label{38}V=\sum_i|\frac{\partial w_{v}}{\partial {\cal
S}_i}|^2+V_{soft}
\end{equation}
where ${\cal S}_i$ denotes the scalar component of the superfields
involved in the model, and $V_{soft}$ includes all possible SUSY
soft terms for the scalar fields ${\cal S}_i$, and it is invariant
under the $T'\otimes Z_3\otimes Z_9$ flavor symmetry.
\begin{equation}
\label{39}V_{soft}=\sum_i m^2_{{\cal S}_i}|{\cal S}_i|^2+...
\end{equation}
where $m^2_{{\cal S}_i}$ is the soft mass, and dots stand for other
soft SUSY breaking bilinear and trilinear operators. By choosing
positive soft mass $m^2_{{\cal S}_i}$ for the driving fields, all
the driving fields don't acquire VEVs. Since the superpotential
$w_v$ is linear in the driving fields, in the SUSY limit all the
derivatives with respect to the scalar components of the superfields
not charged under $U(1)_R$ symmetry vanish. Therefore in discussing
the minimization of the scalar potential, we have to take into
account only the derivatives with respect to the scalar components
of the driving fields, then we have
\begin{eqnarray}
\nonumber \frac{\partial
w_{v}}{\partial\varphi^R_{T1}}&=&ig_1\phi^2_1+g_2\varphi_{T1}\Delta=0\\
\nonumber\frac{\partial
w_{v}}{\partial\varphi^R_{T2}}&=&(1-i)g_1\phi_1\phi_2+g_2\varphi_{T3}\Delta=0\\
\nonumber\frac{\partial
w_{v}}{\partial\varphi^R_{T3}}&=&g_1\phi^2_2+g_2\varphi_{T2}\Delta=0\\
\nonumber\frac{\partial
w_{v}}{\partial\phi^R_{1}}&=&g_3\phi_2\chi+g'_4(\varphi_{T1}\phi_2-(1-i)\varphi_{T3}\phi_1)=0\\
\nonumber\frac{\partial
w_{v}}{\partial\phi^R_{2}}&=&-g_3\phi_1\chi+g'_4(\varphi_{T1}\phi_1+(1+i)\varphi_{T2}\phi_2)=0\\
\nonumber\frac{\partial
w_{v}}{\partial\chi^R}&=&g_5\chi^2+g_6\theta'\theta''+g_7(\varphi^2_{T1}+2\varphi_{T2}\varphi_{T3})=0\\
\nonumber\frac{\partial
w_{v}}{\partial\theta''^R}&=&g_8\theta''^2+g_9\theta'\chi+g_{10}(\varphi^2_{T3}+2\varphi_{T1}\varphi_{T2})=0\\
\nonumber\frac{\partial
w_{v}}{\partial\Delta^R}&=&M_{\Delta}\Delta+g_{11}\chi^2+g_{12}\theta'\theta''+g_{13}(\varphi^2_{T1}+2\varphi_{T2}\varphi_{T3})=0\\
\nonumber\frac{\partial
w_{v}}{\partial\bar{\Delta}^R}&=&\bar{M}_{\Delta}\bar{\Delta}+g_{14}\Delta^2=0\\
\nonumber\frac{\partial
w_{v}}{\partial\varphi^R_{S1}}&=&\frac{2}{3}g_{15}(\varphi^2_{S1}-2\varphi_{S2}\varphi_{S3})+g_{16}\varphi_{S1}\tilde{\xi}=0\\
\nonumber\frac{\partial
w_{v}}{\partial\varphi^R_{S2}}&=&\frac{2}{3}g_{15}(\varphi^2_{S2}-\varphi_{S1}\varphi_{S2})+g_{16}\varphi_{S3}\tilde{\xi}=0\\
\nonumber\frac{\partial
w_{v}}{\partial\varphi^R_{S3}}&=&\frac{2}{3}g_{15}(\varphi^2_{S3}-\varphi_{S1}\varphi_{S2})+g_{16}\varphi_{S2}\tilde{\xi}=0\\
\label{40}\frac{\partial
w_{v}}{\partial\xi^R}&=&g_{17}(\varphi^2_{S1}+2\varphi_{S2}\varphi_{S3})+g_{18}\xi^2+g_{19}\xi\tilde{\xi}+g_{20}\tilde{\xi}^2=0
\end{eqnarray}
where $g'_4=\frac{1-i}{2}g_4$, hereafter we simply denote $g'_4$
with $g_4$ if there is no confusion. These sets of equations admit
the solutions
\begin{eqnarray}
\nonumber~~~~~~~~~~~~~~~ \langle\chi\rangle&=&u_{\chi}\\
\nonumber~~~~~~~~~~~~~~~\langle\theta'\rangle&=&u'_{\theta}=-\Big[\frac{(g^2_3g_7+g^2_4g_5)^2g_8}{g^4_4g^2_6g_9}\Big]^{1/3}u_{\chi}\\
\nonumber~~~~~~~~~~~~~~~\langle\theta''\rangle&=&u''_{\theta}=\Big[\frac{(g^2_3g_7+g^2_4g_5)g_9}{g^2_4g_6g_8}\Big]^{1/3}u_{\chi}\\
\nonumber~~~~~~~~~~~~~~~\langle\Delta\rangle&=&u_{\Delta}=\frac{g^2_3(g_7g_{12}-g_6g_{13})+g^2_4(g_5g_{12}-g_6g_{11})}{g^2_4g_6}\frac{u^2_{\chi}}{M_{\Delta}}\\
\nonumber~~~~~~~~~~~~~~~ \langle\bar{\Delta}\rangle&=&\bar{u}_{\Delta}=-\frac{[g^2_3(g_{7}g_{12}-g_6g_{13})+g^2_4(g_5g_{12}-g_6g_{11})]^2g_{14}}{g^4_4g^2_6}\frac{u^4_{\chi}}{M^2_{\Delta}\bar{M}_{\Delta}}\\
\nonumber~~~~~~~~~~~~~~~\langle\phi\rangle&=&(v_1,0),~~~~~v_1=\Big(\frac{ig_2g_3[g^2_3(g_7g_{12}-g_6g_{13})+g^2_4(g_5g_{12}-g_6g_{11})]}{g_1g^3_4g_6}\Big)^{1/2}M^{\,-1/2}_{\Delta}u^{3/2}_{\chi}\\
\nonumber~~~~~~~~~~~~~~~\langle\varphi_{T}\rangle&=&(v_T,0,0),~~~~~~v_{T}=\frac{g_3}{g_4}u_{\chi}\\
\nonumber~~~~~~~~~~~~~~~\langle\tilde{\xi}\rangle&=&0\\
\nonumber~~~~~~~~~~~~~~~\langle\xi\rangle&=&u_{\xi}\\
\label{41}~~~~~~~~~~~~~~~\langle\varphi_{S}\rangle&=&(v_{S},v_{S},v_{S}),~~~~~v_S=\Big(-\frac{g_{18}}{3g_{17}}\Big)^{1/2}u_{\xi}
\end{eqnarray}
where both $u_{\xi}$ and $u_{\chi}$ are undetermined, by choosing
$m^2_{\xi}$ and $m^2_{\chi}$ to be negative, $u_{\xi}$ and
$u_{\chi}$ would take non-zero values. From Eq.(\ref{41}), we see
that the correct vacuum alignment shown in Eq.(\ref{13}) is
realized. As for the values of the VEVs, we can choose the
parameters in the superpotential $w_{v}$ so that the required orders
of the VEVs in Eq.(\ref{14}) can be achieved.

\section{corrections to the leading order predictions for the fermion masses and flavor mixing}

In the previous section, we have shown that realistic fermion mass
hierarchies and flavor mixing are successfully produced at the
leading order in our model. However, the leading order results would
receive corrections from the higher dimensional operators consistent
with the symmetry of the model, which are suppressed by additional
powers of $\Lambda$. We will study these terms and analyze their
physical effects case by case. The next to leading order corrections
can be classified into two groups: the first class of corrections
are induced by the higher dimensional operators present in the
superpotential $w_{v}$, which can change the vacuum alignment in
Eq.(\ref{13}), therefore the leading order mass matrices are
modified. The second are induced by the higher dimensional operators
in the Yukawa superpotentials $w_{\ell}$ and $w_{q}$, which could
modify the Yukawa couplings after the electroweak and flavor
symmetry breaking .

\subsection{Higher dimensional operators in the flavon superpotential and the corrections to the vacuum alignment}
If we include the next to leading order operators in the flavon
superpotential $w_{v}$, the vacuum alignment in Eq.(\ref{13}) would
be modified, and the higher order corrections to the vacuum
alignment are discussed in detail in the Appendix B. The corrections
result in a shift in the VEVs of the scalar fields, and therefore
the new vacuum configuration is given by
\begin{eqnarray}
\nonumber&&\langle\varphi_{T}\rangle=(v_T+\delta v_{T1},\delta
v_{T2},\delta v_{T3}),~~~\langle\varphi_S\rangle=(v_S+\delta v_{S1},v_S+\delta v_{S2},v_S+\delta v_{S3}),\\
\nonumber&&\langle\phi\rangle=(v_1+\delta v_1,\delta
v_2),~~~\langle\xi\rangle=u_{\xi},~~~\langle\tilde{\xi}\rangle=\delta\tilde{u}_{\xi},~~~\langle\theta'\rangle=u'_{\theta}+\delta
u'_{\theta},\\
\label{42}&&\langle\theta''\rangle=u''_{\theta}+\delta
u''_{\theta},~~~\langle\Delta\rangle=u_{\Delta}+\delta
u_{\Delta},~~~\langle\bar{\Delta}\rangle=\bar{u}_{\Delta}+\delta\bar{u}_{\Delta},~~~\langle\chi\rangle=u_{\chi}
\end{eqnarray}
In the Appendix B, we show that the corrections $\delta v_{T2}$,
$\delta v_{T3}$, $\delta v_{1}$, $\delta v_{2}$, $\delta
u'_{\theta}$, $\delta u''_{\theta}$ and $\delta\bar{u}_{\Delta}$
arise at order $1/\Lambda$. $\delta v_{T1}$ and $\delta u_{\Delta}$
are of order $1/\Lambda^2$, and the corrections $\delta v_{S1}$,
$\delta v_{S2}$, $\delta v_{S3}$ and $\delta \tilde{u}_{\xi}$ are
suppressed by $1/\Lambda^3$, which are small enough and can be
negligible. Note that there should also be corrections to the VEVs
of $\xi$ and $\chi$, but we do not have to indicate this explicitly
by the addition of terms $\delta u_{\xi}$ and $\delta u_{\chi}$,
since both $u_{\xi}$ and $u_{\chi}$ are undetermined at the leading
order.

Repeating the calculations in section IV and substituting the
modified vacuum into the Yukawa superpotentials $w_{\ell}$ and
$w_{q}$, we can obtain the new vacuum corrections to the fermion
mass matrices as follows
\begin{eqnarray}
\label{43}\delta M^e_1&=&\left(\begin{array}{ccc}
y_{e}\frac{\bar{u}^2_{\Delta}\delta v_{T1}+2\bar{u}_{\Delta}\delta\bar{u}_{\Delta}v_T}{\Lambda^3}&y_{e}\frac{\bar{u}^2_{\Delta}\delta v_{T3}}{\Lambda^3}&y_{e}\frac{\bar{u}^2_{\Delta}\delta v_{T2}}{\Lambda^3}\\
y_{\mu2}\frac{u_{\Delta}\delta v_{T2}}{\Lambda^2}&2iy_{\mu1}\frac{v_1\delta v_1}{\Lambda^2}&(1-i)y_{\mu1}\frac{v_1\delta v_2}{\Lambda^2}+y_{\mu2}\frac{u_{\Delta}\delta v_{T3}}{\Lambda^2}\\
y_{\tau}\frac{\delta v_{T3}}{\Lambda}&y_{\tau}\frac{\delta
v_{T2}}{\Lambda}&y_{\tau}\frac{\delta v_{T1}}{\Lambda}
\end{array}\right)v_d\\
\label{44}\delta M^{\nu}_1&=&\left(\begin{array}{ccc}
\frac{4}{3}y_S\frac{\delta v_{S1}}{\Lambda}&-\frac{2}{3}y_S\frac{\delta v_{S3}}{\Lambda}&-\frac{2}{3}y_S\frac{\delta v_{S2}}{\Lambda}\\
-\frac{2}{3}y_S\frac{\delta
v_{S3}}{\Lambda}&\frac{4}{3}y_S\frac{\delta
v_{S2}}{\Lambda}&-\frac{2}{3}y_{S}\frac{\delta
v_{S1}}{\Lambda}\\
-\frac{2}{3}y_S\frac{\delta
v_{S2}}{\Lambda}&-\frac{2}{3}y_{S}\frac{\delta
v_{S1}}{\Lambda}&\frac{4}{3}y_S\frac{\delta v_{S3}}{\Lambda}
\end{array}\right)\frac{v^2_u}{\Lambda}\\
\label{45}\delta M^{u}_1&=&\left(\begin{array}{ccc}
iy_{u1}\frac{u_{\Delta}\delta v_{T2}}{\Lambda^2}&\delta u_1&-y_{u5}\frac{\delta v_2}{\Lambda}\\
\delta u'_1&y_{u1}\frac{u_{\Delta}\delta v_{T1}+\delta u_{\Delta}v_T}{\Lambda^2}+2iy_{u2}\frac{v_1\delta v_1}{\Lambda^2}&y_{u5}\frac{\delta v_1}{\Lambda}\\
-y_{u4}\frac{\delta v_2}{\Lambda}&y_{u4}\frac{\delta v_1}{\Lambda}&0
\end{array} \right)v_u\\
\label{46}\delta M^{d}_1&=&\left(\begin{array}{ccc}
iy_{d1}\frac{\bar{u}_{\Delta}\delta v_{T2}}{\Lambda^2}&\delta d_1&-y_{d4}\frac{u_{\Delta}\delta v_2}{\Lambda^2}\\
\delta d'_1&y_{d1}\frac{\bar{u}_{\Delta}\delta v_{T1}+\delta\bar{u}_{\Delta}v_T}{\Lambda^2}&y_{d4}\frac{u_{\Delta}\delta v_1+\delta u_{\Delta}v_1}{\Lambda^2}\\
-y_{d3}\frac{u_{\Delta}\delta
v_2}{\Lambda^2}&y_{d3}\frac{u_{\Delta}\delta v_1+\delta
u_{\Delta}v_1}{\Lambda^2}&y_b\frac{\delta u_{\Delta}}{\Lambda}
\end{array}\right)v_d
\end{eqnarray}
where $\delta e_1$, $\delta e'_1$, $\delta u_1$, $\delta u'_1$,
$\delta d_1$ and $\delta d'_1$ are given by
\begin{eqnarray}
\nonumber\delta u_1&=&\frac{1-i}{2}y_{u1}\frac{u_{\Delta}\delta
v_{T3}}{\Lambda^2}-iy_{u2}\frac{v_1\delta
v_2}{\Lambda^2}-y_{u3}\frac{\delta
u''_{\theta}u_{\Delta}+u''_{\theta}\delta u_{\Delta}}{\Lambda^2}\\
\nonumber\delta u'_1&=&\frac{1-i}{2}y_{u1}\frac{u_{\Delta}\delta
v_{T3}}{\Lambda^2}-iy_{u2}\frac{v_1\delta
v_2}{\Lambda^2}+y_{u3}\frac{\delta
u''_{\theta}u_{\Delta}+u''_{\theta}\delta u_{\Delta}}{\Lambda^2}\\
\nonumber\delta
d_1&=&\frac{1-i}{2}y_{d1}\frac{\bar{u}_{\Delta}\delta
v_{T3}}{\Lambda^2}-y_{d2}\frac{\delta
u''_{\theta}\bar{u}_{\Delta}+u''_{\theta}\delta\bar{u}_{\Delta}}{\Lambda^2}\\
\nonumber\delta
d'_1&=&\frac{1-i}{2}y_{d1}\frac{\bar{u}_{\Delta}\delta
v_{T3}}{\Lambda^2}+y_{d2}\frac{\delta
u''_{\theta}\bar{u}_{\Delta}+u''_{\theta}\delta\bar{u}_{\Delta}}{\Lambda^2}
\end{eqnarray}
As for $\delta M^{e}_1$, we have neglected the corrections induced
by $\delta v_{Si}(i=1,2,3)$ and $\delta \tilde{u}_{\xi}$, since they
are of higher order $1/\Lambda^3$ and can be negligible comparing
with the corrections proportional to $\delta v_{Ti}(i=1,2,3)$ and
$\delta\bar{u}_{\Delta}$. Eq.(\ref{44}) implies that the corrections
to the neutrino mass matrix are suppressed by additional power of
$1/\Lambda^3$ relative to the leading order results. Concerning the
quark sector, the correction terms $y_{u3}\frac{\delta
u''_{\theta}u_{\Delta}+u''_{\theta}\delta u_{\Delta}}{\Lambda^2}$,
$y_{d2}\frac{\delta
u''_{\theta}\bar{u}_{\Delta}+u''_{\theta}\delta\bar{u}_{\Delta}}{\Lambda^2}$
and $y_b\frac{\delta u_{\Delta}}{\Lambda}$ can be absorbed by the
redefinition of $y_{u3}$, $y_{d2}$ and $y_b$ respectively.

\subsection{Corrections induced by higher dimensional operators in the Yukawa superpotential }
\begin{enumerate}
\item{Corrections to $w_{\ell}$}

The leading order operators relevant to $e^{c}$ are of order
$1/\Lambda^3$, which are shown in Eq.(\ref{16}), at the next order
$1/\Lambda^4$ there are two operators
\begin{equation}
\label{47}
e^{c}(\ell\varphi_T)\Delta^2\bar{\Delta}H_{d}/\Lambda^4,~~~~~~~e^{c}(\ell\phi\phi)\Delta\bar{\Delta}H_{d}/\Lambda^4
\end{equation}
Because
$(\phi\phi)_{\mathbf{3}}=(i\phi^2_1,\phi^2_2,(1-i)\phi_1\phi_2)$,
its VEV is parallel to that of $\varphi_T$. Therefore both operators
have the same structure as the leading operator
$e^{c}(\ell\varphi_T)\bar{\Delta}^2H_{d}/\Lambda^3$, their effects
can be absorbed by the redefinition of $y_e$. Concerning the
$\mu^{c}$ relevant terms in the leading order Yukawa superpotential
in Eq.(\ref{16}), they comprise both terms of order $1/\Lambda^2$
and terms of order $1/\Lambda^3$. The subleading operators invariant
under the symmetry of the model arise at order $1/\Lambda^5$, and
their contributions are completely negligible relative to the
corrections from the modified vacuum configuration. The leading
operator of the $\tau^c$ relevant term is of order $1/\Lambda$, and
the next to leading order corrections are of order $1/\Lambda^4$,
therefore their contributions can be neglected comparing with the
corrections from the new vacuum.

The same arguments used for the charged lepton mass are applicable
to the neutrino sector as well. The leading operators contributing
to $M^{\nu}$ are of order $1/\Lambda^2$ from Eq.(\ref{23}), and the
leading order results receive corrections from higher dimensional
operators of order $1/\Lambda^5$ at the next to leading order.
Therefore, the charged lepton mass matrix mainly receives
corrections from the modified vacuum configuration. Whereas, the
corrections to the neutrino mass matrix from the next to leading
order operators in both $w_{v}$ and $w_{\nu}$ are negligible, and
$T'$ is approximately broken to $Z_4$ subgroup in the neutrino
sector, even if higher order corrections are included.
\item{Corrections to $w_{u}$}

As is shown in Eq.(\ref{30}), the leading operators, which give rise
to the $M^{u}_{11}$, $M^{u}_{12}$, $M^{u}_{21}$ and $M^{u}_{22}$,
are of order $1/\Lambda^2$. At the next order $1/\Lambda^3$, there
are two operators whose contributions can not be absorbed by
parameter redefinition
\begin{equation}
\label{49}x_{u1}((Q_LU^c)_{\mathbf{3}}(\varphi_T\varphi_T)_{\mathbf{3}_S})'\theta''H_{u}/\Lambda^3,~~~~x_{u2}((Q_LU^c)_{\mathbf{3}}(\varphi_T\varphi_T)_{\mathbf{3}_S})''\theta'H_{u}/\Lambda^3
\end{equation}
The leading operators contributing to $M^{u}_{13}$, $M^{u}_{23}$,
$M^{u}_{31}$ and $M^{u}_{32}$ are of order $1/\Lambda$ from
Eq.(\ref{30}), and the next to leading order corrections arise at
order $1/\Lambda^4$. These contributions are negligible relative to
the corrections induced by the modified vacuum, which is shown in
Eq.(44). The corrections to $M^{u}_{33}$ can be absorbed by
redefining the parameter $y_t$. As a result, the higher dimensional
operators corrections to the up quark mass matrix are
\begin{equation}
\label{50}\delta M^{u}_2=\left(\begin{array}{ccc}
\frac{2i}{3}x_{u2}\frac{u'_{\theta}v^2_{T}}{\Lambda^3}&\frac{1-i}{3}x_{u1}\frac{u''_{\theta}v^2_T}{\Lambda^3}&0\\
\frac{1-i}{3}x_{u1}\frac{u''_{\theta}v^2_T}{\Lambda^3}&0&0\\
0&0&0
\end{array} \right)v_{u}
\end{equation}
\item{Corrections to $w_d$}

Concerning the $M^{d}_{11}$, $M^{d}_{12}$,$M^{d}_{21}$ and
$M^{d}_{22}$ relevant operators, the leading terms are of order
$1/\Lambda^2$, which are shown in Eq.(\ref{31}), and there are three
operators at the order $1/\Lambda^3$
\begin{equation}
\label{51}(\varphi_{T}Q_LD^c)\Delta^2H_{d}/\Lambda^3,~~~~((Q_LD^c)_{\mathbf{3}}(\phi\phi)_{\mathbf{3}})\Delta
H_{d}/\Lambda^3,~~~~(Q_LD^c)'\theta''\Delta^2H_{d}/\Lambda^3
\end{equation}
After electroweak and flavor symmetry breaking, the above operators
have the same structures as the leading ones, and their
contributions can be absorbed by redefinition of $y_{d1}$ and
$y_{d2}$. Nontrivial higher dimensional operators arise at the order
$1/\Lambda^4$, their contributions are negligible comparing with the
corrections from the modified vacuum, which is shown in Eq.(45).
Similarly the next to leading operators contributing to
$M^{d}_{13}$, $M^{d}_{23}$, $M^{d}_{31}$ and $M^{d}_{32}$ are of
order $1/\Lambda^3$, and only two operators remain after symmetry
breaking and parameter redefinition
\begin{equation}
\label{52}
x_{d1}(\varphi_TQ_L\phi)b^{c}\theta''H_{d}/\Lambda^3,~~~~x_{d2}Q_3(\varphi_TD^{c}\phi)''\theta''H_{d}/\Lambda^3
\end{equation}
Therefore the corrections to the down quark mass matrix from the
higher dimensional operators are
\begin{equation}
\label{53}\delta M^{d}_{2}=\left(\begin{array}{ccc}
0&0&ix_{d2}\frac{u''_{\theta}v_1v_{T}}{\Lambda^3}\\
0&0&0\\
ix_{d1}\frac{u''_{\theta}v_1v_{T}}{\Lambda^3}&0&0
\end{array}\right)v_d
\end{equation}
\end{enumerate}
\subsection{Fermion masses and flavor mixing including the next to leading order corrections}

\begin{enumerate}
\item{Lepton masses and MNSP matrix}

Combining the leading order predictions Eq.(\ref{18}) for the
charged lepton mass matrix with the subleading corrections in
Eq.(\ref{43}), we obtain that the charged lepton mass matrix is
modified as
\begin{equation}
\label{54}{{\cal M}^{e}}=M^{e}+\delta
M^{e}_1=\left(\begin{array}{ccc}
y_{e}\frac{\bar{u}^2_{\Delta}v_T}{\Lambda^3}+y'_e\frac{v^3_S}{\Lambda^3}&y_e\frac{\bar{u}^2_{\Delta}\delta v_{T3}}{\Lambda^3}+y'_e\frac{v^{3}_S}{\Lambda^3} &y_e\frac{\bar{u}^2_{\Delta}\delta v_{T2}}{\Lambda^3}+y'_e\frac{v^3_S}{\Lambda^3}\\
\delta e'_2&y_{\mu}\frac{v^2_1}{\Lambda^2} &\delta e_2\\
y_{\tau}\frac{\delta v_{T3}}{\Lambda}& y_{\tau}\frac{\delta
v_{T2}}{\Lambda}& y_{\tau}\frac{v_{T}}{\Lambda}
\end{array}\right)
\end{equation}
where
\begin{eqnarray}
\nonumber\delta e_2&=&(1-i)y_{\mu1}\frac{v_1\delta
v_2}{\Lambda^2}+y_{\mu2}\frac{u_{\Delta}\delta
v_{T3}}{\Lambda^2}+y_{\mu\tau}\frac{u''_{\theta}v^2_T}{\Lambda^3}\\
\nonumber\delta e'_2&=&y_{\mu2}\frac{u_{\Delta}\delta
v_{T2}}{\Lambda^2}+y_{\mu e}\frac{u'_{\theta}v^2_T}{\Lambda^3}
\end{eqnarray}
we have set $v_T+\delta v_{T1}\rightarrow v_{T}$, $v_1+\delta
v_1\rightarrow v_1$ and
$\bar{u}_{\Delta}+\delta\bar{u}_{\Delta}\rightarrow\bar{u}_{\Delta}$.
In the neutrino sector, since the corrections from the new vacuum
and the higher dimensional operators in the the Yukawa
superpotential $w_{\nu}$ are of order $1/\Lambda^5$, as are shown in
the previous subsections, these contributions are negligible.
Therefore the neutrino mass matrix is approximately not affected by
the subleading operators. Performing the same procedure as that in
section IV, we see that both the charged lepton masses and the
neutrino masses approximately are not modified by the next to
leading order operators, and the MNSP matrix becomes
\begin{equation}
\label{55}V_{\rm MNSP}\approx\left(\begin{array}{ccc}
\sqrt{\frac{2}{3}}+\frac{1}{\sqrt{6}}\frac{\delta v^*_{T3}}{v^*_{T}}&\frac{1}{\sqrt{3}}-\frac{1}{\sqrt{3}}\frac{\delta v^*_{T3}}{v^*_{T}}&-\frac{1}{\sqrt{2}}\frac{\delta v^*_{T3}}{v^*_{T}}\\
-\frac{1}{\sqrt{6}}+\frac{1}{\sqrt{6}}\frac{\delta v^*_{T2}}{v^*_T}&\frac{1}{\sqrt{3}}-\frac{1}{\sqrt{3}}\frac{\delta v^*_{T2}}{v^*_T}&-\frac{1}{\sqrt{2}}-\frac{1}{\sqrt{2}}\frac{\delta v^*_{T2}}{v^*_T}\\
-\frac{1}{\sqrt{6}}+\frac{1}{\sqrt{6}}\frac{2\delta v_{T3}-\delta
v_{T2}}{v_T}&\frac{1}{\sqrt{3}}+\frac{1}{\sqrt{3}}\frac{\delta
v_{T2}+\delta
v_{T3}}{v_T}&\frac{1}{\sqrt{2}}-\frac{1}{\sqrt{2}}\frac{\delta
v_{T2}}{v_T}
\end{array}
\right)
\end{equation}
therefore
\begin{eqnarray}
\nonumber&&|(V_{\rm MNSP})_{e3}|\approx\frac{1}{\sqrt{2}}\Big|\frac{\delta v_{T3}}{v_{T}}\Big|\\
\nonumber&&\tan^2\theta_{23}\approx1+2\Big(\frac{\delta
v_{T2}}{v_{T}}+\frac{\delta v^*_{T2}}{v^*_{T}}\Big)\\
\label{56}&&\tan^2\theta_{12}\approx\frac{1}{2}-\frac{3}{4}\Big(\frac{\delta
v_{T3}}{v_{T}}+\frac{\delta v^*_{T3}}{v^*_{T}}\Big)
\end{eqnarray}
We see $\delta v_{T2}/v_{T}\sim\lambda^2$ and $\delta
v_{T3}/v_{T}\sim\lambda^2$ from the Appendix B, therefore the
deviations of the mixing angles from the TB mixing predictions are
of order $\lambda^2$, which are allowed by the current neutrino
oscillation data in Eq.(\ref{8}).
\item{Quark masses and CKM matrix}

Including the corrections $\delta M^{u}_i$ and $\delta M^{d}_i
(i=1,2)$ induced by the new vacuum and the higher dimensional
operators in the Yukawa superpotential $w_q$, the up quark and down
quark mass matrices becomes
\begin{eqnarray}
\nonumber{\cal M}^{u}&=&M^{u}+\delta M^{u}_1+\delta
M^{u}_2=\left(\begin{array}{ccc} iy_{u1}\frac{u_{\Delta}\delta
v_{T2}}{\Lambda^2}+\frac{2i}{3}x_{u2}\frac{u'_{\theta}v^2_T}{\Lambda^3}&-y_{u3}\frac{u''_{\theta}u_{\Delta}}{\Lambda^2}+\delta u_2&-y_{u5}\frac{\delta v_2}{\Lambda}\\
y_{u3}\frac{u''_{\theta}u_{\Delta}}{\Lambda^2}+\delta u_2&y_{u1}\frac{u_{\Delta}v_{T}}{\Lambda^2}+iy_{u2}\frac{v^2_1}{\Lambda^2}&y_{u5}\frac{v_1}{\Lambda}\\
-y_{u4}\frac{\delta v_2}{\Lambda}&y_{u4}\frac{v_1}{\Lambda}&y_t
\end{array}\right)v_u\\
\nonumber{\cal M}^{d}&=&\left(\begin{array}{ccc}
iy_{d1}\frac{\bar{u}_{\Delta}\delta v_{T2}}{\Lambda^2}&-y_{d2}\frac{u''_{\theta}\bar{u}_{\Delta}}{\Lambda^2}+\frac{1-i}{2}y_{d1}\frac{\bar{u}_{\Delta}\delta v_{T3}}{\Lambda^2}&-y_{d4}\frac{u_{\Delta}\delta v_2}{\Lambda^2}+ix_{d2}\frac{u''_{\theta}v_1v_T}{\Lambda^3}\\
y_{d2}\frac{u''_{\theta}\bar{u}_{\Delta}}{\Lambda^2}+\frac{1-i}{2}y_{d1}\frac{\bar{u}_{\Delta}\delta v_{T3}}{\Lambda^2}&y_{d1}\frac{\bar{u}_{\Delta}v_T}{\Lambda^2}&y_{d4}\frac{u_{\Delta}v_1}{\Lambda^2}\\
-y_{d3}\frac{u_{\Delta}\delta
v_2}{\Lambda^2}+ix_{d1}\frac{u''_{\theta}v_1v_T}{\Lambda^3}&y_{d3}\frac{u_{\Delta}v_1}{\Lambda^2}&y_b\frac{u_{\Delta}}{\Lambda}
\end{array}\right)v_d
\end{eqnarray}
where $\delta u_2=\frac{1-i}{2}y_{u1}\frac{u_{\Delta}\delta
v_{T3}}{\Lambda^2}-iy_{u2}\frac{v_1\delta
v_2}{\Lambda^2}+\frac{1-i}{3}x_{u1}\frac{u''_{\theta}v^2_T}{\Lambda^3}$,
and we have set $v_T+\delta v_{T1}\rightarrow v_{T}$, $v_1+\delta
v_1\rightarrow v_1$, $u_{\Delta}+\delta u_{\Delta}\rightarrow
u_{\Delta}$ and $\bar{u}_{\Delta}+\delta\bar{u}_{\Delta}\rightarrow
\bar{u}_{\Delta}$. Diagonalizing the above mass matrices
perturbatively, we obtain the quark masses as follows
\begin{eqnarray}
\nonumber m_{u}&\approx&\Big|\Big(iy_{u1}\frac{u_{\Delta}\delta
v_{T2}}{\Lambda^2}-iy_{u2}\frac{\delta
v^2_2}{\Lambda^2}+\frac{y^2_{u3}y_tu''^2_{\theta}u^2_{\Delta}}{(iy_{u2}y_t-y_{u4}y_{u5})v^2_1\Lambda^2}+\frac{2i}{3}x_{u2}\frac{u'_{\theta}v^2_T}{\Lambda^3}\Big)v_u\Big|\\
\nonumber
m_{c}&\approx&\Big|\Big(iy_{u2}-\frac{y_{u4}y_{u5}}{y_t}\Big)\frac{v^2_1}{\Lambda^2}v_u\Big|\\
\nonumber m_{t}&\approx&\Big|y_tv_u\Big|\\
\nonumber
m_{d}&\approx&\Big|\Big(iy_{d1}\frac{\bar{u}_{\Delta}\delta
v_{T2}}{\Lambda^2}+\frac{y^2_{d2}}{y_{d1}}\frac{u''^2_{\theta}\bar{u}_{\Delta}}{v_T\Lambda^2}\Big)v_d\Big|\\
\nonumber
m_s&\approx&\Big|y_{d1}\frac{\bar{u}_{\Delta}v_T}{\Lambda^2}v_d\Big|\\
\label{57}m_b&\approx&\Big|y_b\frac{u_{\Delta}}{\Lambda}v_d\Big|
\end{eqnarray}
and the CKM matrix elements are approximately given by
\begin{eqnarray}
\nonumber V_{ud}&\approx& V_{cs}\approx V_{tb}\approx1\\
\nonumber
V^{*}_{us}&\approx&-V_{cd}\approx\frac{y_{d2}}{y_{d1}}\frac{u''_{\theta}}{v_{T}}+\frac{1-i}{2}\frac{\delta
v_{T3}}{v_T}+\frac{\delta
v_2}{v_1}-\frac{y_{u3}y_tu''_{\theta}u_{\Delta}}{(iy_{u2}y_t-y_{u4}y_{u5})v^2_1}\\
\nonumber V^{*}_{cb}&\approx&-V_{ts}\approx(\frac{y_{d3}}{y_b}-\frac{y_{u4}}{y_t})\frac{v_1}{\Lambda}\\
\nonumber V^{*}_{ub}&\approx&-\frac{y_{u3}y_t}{iy_{u2}y_t-y_{u4}y_{u5}}(\frac{y_{d3}}{y_b}-\frac{y_{u4}}{y_t})\frac{u''_{\theta}u_{\Delta}}{v_1\Lambda}+i\frac{x_{d1}}{y_b}\frac{u''_{\theta}v_1v_{T}}{u_{\Delta}\Lambda^2}\\
\label{58}V_{td}&\approx&\frac{y_{d2}}{y_{d1}}(\frac{y_{d3}}{y_b}-\frac{y_{u4}}{y_t})\frac{u''_{\theta}v_1}{v_T\Lambda}+(\frac{y_{d3}}{y_b}-\frac{y_{u4}}{y_t})(\frac{\delta
v_2}{\Lambda}+\frac{1-i}{2}\frac{v_1\delta
v_{T3}}{v_T\Lambda})-i\frac{x_{d1}}{y_b}\frac{u''_{\theta}v_1v_T}{u_{\Delta}\Lambda^2}
\end{eqnarray}
Because $\delta v_{T2}/\Lambda$, $\delta v_{T3}/\Lambda$ and $\delta
v_2/\Lambda$ are of order $\lambda^4$ from the Appendix B, to get
the appropriate magnitude of the up quark mass, we assume that the
couplings $y_{u1}$ and $x_{u2}$ are smaller than one by a factor of
$\lambda$, i.e. $y_{u1}\sim x_{u2}\sim\lambda$. From Eq.(\ref{14}),
Eq.(\ref{57}) and Eq.(\ref{58}), we see that the realistic
hierarchies in quark masses and CKM matrix elements are generated,
and the relations between quark masses and mixing angles in
Eq.(\ref{36}) are no longer satisfied after including the subleading
contributions. It is very likely that the higher order contributions
would improve the agreement between the model predictions and the
experimental data.
\end{enumerate}

\section{summary and discussion }

$T'$ is a promising discrete group for a unified description of both
quark and lepton mass hierarchies and flavor mixing. $T'$ can
reproduce the success of $A_4$ model building, and $T'$ has
advantage over $A_4$ in extension to the quark sector because it has
doublet representations in addition to singlet representations and
triplet representation. We have built a SUSY model based on
$T'\otimes Z_3\otimes Z_9$ flavor symmetry, where the fermion mass
hierarchies arise from the flavor symmetry breaking which is crucial
in producing the flavor mixing as well.

In the lepton sector, the left handed electroweak lepton doublets
$l_i(i=1,2,3)$ are $T'$ triplet, and the right handed charged
leptons $e^{c}$, $\mu^{c}$ and $\tau^{c}$ transform as $\mathbf{1}$,
$\mathbf{1}''$ and $\mathbf{1}'$ respectively. The charged lepton
mass matrix is no longer diagonal at the leading order, and $T'$ is
broken completely in the charged lepton sector. However, it is
broken down to the $Z_3$ subgroup generated by the element $T$ in
$A_4$ model\cite{Altarelli:2005yx} and in the $T'$ model of
Ref.\cite{Feruglio:2007uu} at the leading order, then it is further
broken to nothing by the subleading operators. The MNSP matrix is
predicted to be nearly TB mixing matrix at the leading order, and
the deviations due to the contributions of the charged lepton sector
are of order $\lambda^3$ and are negligible. In the neutrino sector,
$T'$ is broken down to the $Z_4$ subgroup generated by the element
$TST^2$ at the leading order as Ref.\cite{Feruglio:2007uu}. The
higher order corrections to the neutrino mass matrix are strongly
suppressed, consequently the $Z_4$ symmetry almost remains.
Considering the next to leading order operators in the Yukawa
superpotential $w_{\ell}$ and the flavon superpotential $w_v$, then
the mixing angles are predicted to deviate from the TB mixing
predictions by terms of order $\lambda^2$, which are in the interval
indicated by the experimental data at the $3\sigma$ level.

In the quark sector, doublet representation are exploited. The first
two generations transform as doublet ($\mathbf{2}$ or
$\mathbf{2}'$), and the third generation is $T'$ singlet
($\mathbf{1}'$ or $\mathbf{1}''$). At the leading order, both the up
and down quark Yukawa matrices textures in U(2) flavor theory are
produced, and the correct hierarchies in quark masses and mixing
angles are obtained. $T'$ is completely broken at the leading order,
this is in contrast with Ref.\cite{Feruglio:2007uu}, where the $T'$
flavor symmetry is broken down to $Z_3$ at the leading order and is
further broken to nothing by the next to leading order
contributions. After including the corrections induced by the next
to leading order operators, the correct orders of quark masses and
CKM matrix elements at the leading order remain except the up quark
mass, we need to mildly fine-tune the coupling coefficients $y_{u1}$
and $x_{u2}$ to be smaller than one by a factor of $\lambda$.

The vacuum alignment and the higher order corrections are discussed
in details. We have shown that the scalar potential in the model
presents the correct $T'$ breaking alignment in a finite portion of
the parameter space, which plays an important role in producing the
realistic fermion mass hierarchies and flavor mixing. The VEVs
should be of the orders shown in Eq.(\ref{14}), the minor hierarchy
in the VEVs can be achieved by moderately fine-tuning the parameters
in $w_v$. The origin of this hierarchies may be qualitatively
understood in the grand unification models\cite{ding}, in which
$b-\tau$ unification may be predicted as well. The higher order
corrections are due to the higher dimensional operators which modify
the Yukawa couplings and the the vacuum alignment, and they don't
spoil the leading order predictions.

Our model is different from the model in Ref.\cite{Feruglio:2007uu}
mainly in the following three aspects:
\begin{enumerate}
\item {We have introduced the auxiliary discrete symmetry $Z_9$ instead of
continuous $U(1)_{FN}$, both the fermion mass hierarchies and flavor
mixing arise from the spontaneous breaking of the flavor symmetry,
whereas a continuous abelian flavor symmetry $U(1)_{FN}$ is
introduced to describe the fermion mass hierarchies in
Ref.\cite{Feruglio:2007uu}.}

\item{There are large differences between the
two models in the quark sector, at the leading order, the favorable
Yukawa matrix textures of the U(2) flavor theory are obtained, and
the realistic quark mass hierarchies and CKM matrix elements are
produced in our model. However, in the model in
Ref.\cite{Feruglio:2007uu}, only the masses of the second and third
generation quarks and the mixing between them are generated at the
leading order, the masses and mixing angles of the first generation
quarks are produced via subleading effects induced by the higher
dimensional operators.}

\item{The large mass hierarchy
between the top quark and the bottom quark arises from the flavor
symmetry breaking, and $\tan\beta\equiv v_{u}/v_{d}$ is of order one
in our model. Nevertheless, this hierarchy is due to large
$\tan\beta$ in Ref.\cite{Feruglio:2007uu}, therefore the quark
masses and mixing angles may receive large radiative corrections,
and the successful predictions in Ref.\cite{Feruglio:2007uu} could
be destroyed at low energy.}
\end{enumerate}
We would like to stress that the origin of all the above differences
is due to the different flavor symmetry ( discrete $Z_9$ instead of
continuous $U(1)_{FN}$ ), different charge assignments and different
flavor symmetry breaking patterns.

As most flavor models, there are a large number of operators with
dimensionless order one coefficients in front of them, However,
experimental tests of this model is not impossible\cite{ding}. Since
quarks and leptons have their superpartners in SUSY, the flavor
symmetry would affect mass matrices of squarks and sleptons as well,
and specific pattern of sfermion masses would be predicted, which
could be tested by measuring squark and slepton masses in future
experiments. Moreover, the squarks and sleptons mass matrices are
severely constrained by the experiments of flavor changing neutral
current (FCNC) processes, and the off-diagonal elements of sfermion
mass matrices must be suppressed in the super-CKM basis. Hence
searching for FCNC and CP violating phenomena such as lepton flavor
violation $\mu\rightarrow e\gamma$ and $\mu-e$ conversion in atom,
electric dipole moments of the electron and neutron, and $B-\bar{B}$
mixing etc are also possible tests of the model. Moreover, we should
check whether there is some accidental continuous symmetry in the
scalar potential, which could affect the above FCNC
processes\cite{Bazzocchi:2007na,Bazzocchi:2004dw}. In addition the
cosmological consequences of the $T'$ flavor symmetry and its
spontaneous broken deserve studying further\cite{cosmology}.

The Yukawa superpotential consists of non-renormalizable
interactions except the top quark relevant term in the model, and a
lot of non-renormalizable operators are involved in the higher order
corrections. These non-renormalizable interactions may be generated
from a renormalizable theory by integrating out some heavy
fields\cite{ding}. Searching for the origin of these
non-renormalizable interactions is a challenging and interesting
question, then the free parameters of the model would be drastically
reduced, and the model become more predictive.
\section*{ACKNOWLEDGEMENTS}
\indent I am grateful to Prof. Dao-Neng Gao and Mu-Lin Yan for very
helpful and stimulating discussions. This work is supported by the
China Postdoctoral Science foundation (20070420735).

\begin{appendix}
\section{basic properties of the discrete group $T'$}
The group $T'$ is denoted as $24/13$ in the Thomas-Wood
classification\cite{group}, and it is isomorphic to the group
$SL_2(F_3)$\cite{Carr:2007qw,Frampton:2007et}, which consists of
$2\times2$ unimodular matrices whose elements are added and
multiplied as integers modulo 3. $T'$ is the double cover of $A_4$
which is the even permutation of 4 objects, and the order of $T'$ is
24. Geometrically, $T'$ is proper rotations leaving a regular
tetrahedron invariant in the SU(2) space. $T'$ can be generated by
two generators $S$ and $T$ with the multiplication
rules\cite{qtp,group}
\begin{equation}
\label{a1}S^4=T^3=1,~~~TS^2=S^2T,~~~ST^{-1}S=TST
\end{equation}
The 24 elements can be written in the form $T^{l}S^{m}T^{n}$, where
$l=0,\pm1$, and if $m=0$ or 2 then $n=0$, while if $m=\pm1$ then
$n=0,\pm1$.

The character table, the explicit matrix representations and the
Clebsch-Gordan coefficients of $T'$ have already been
calculated\cite{qtp}, which are reformulated in
Ref.\cite{Feruglio:2007uu}. $T'$ has seven inequivalent irreducible
representations: three singlet representations $\mathbf{1}^{0}$ and
$\mathbf{1}^{\pm}$, three doublet representations $\mathbf{2}^{0}$
and $\mathbf{2}^{\pm}$, and one triplet representation $\mathbf{3}$.
The triality superscript can describe the multiplication rules of
these representations concisely. We identify $\pm$ as $\pm1$,
trialities add under addition modulo three, and the multiplication
rules are as follows
\begin{eqnarray}
\nonumber&&
\mathbf{1}^{i}\otimes\mathbf{1}^{j}=\mathbf{1}^{i+j}\;,~~~\mathbf{1}^{i}\otimes\mathbf{2}^{j}=\mathbf{2}^{j}\otimes\mathbf{1}^{i}=\mathbf{2}^{i+j}~~({\rm{with}}~ i,j=0,\pm1)\\
\nonumber&&\mathbf{1}^{i}\otimes\mathbf{3}=\mathbf{3}\otimes\mathbf{1}^{i}=\mathbf{3}\;,~~~\mathbf{2}^{i}\otimes\mathbf{2}^{j}=\mathbf{3}\oplus\mathbf{1}^{i+j}\\
\label{a2}&&\mathbf{2}^{i}\otimes\mathbf{3}=\mathbf{3}\otimes\mathbf{2}^{i}=\mathbf{2}^{0}\oplus\mathbf{2}^{+}\oplus\mathbf{2}^{-}\;,\mathbf{3}\otimes\mathbf{3}=\mathbf{3}_{S}\oplus\mathbf{3}_{A}\oplus\mathbf{1}^{0}\oplus\mathbf{1}^{+}\oplus\mathbf{1}^{-}
\end{eqnarray}
where the triality notations are related the usually used notations
$\mathbf{1}$, $\mathbf{1}'$, $\mathbf{1}''$, $\mathbf{2}$,
$\mathbf{2}'$ and $\mathbf{2}''$
\cite{Ma:2001dn,Babu:2002dz,Ma:2004zv,Altarelli:2005yp,Chen:2005jm,Zee:2005ut,Altarelli:2005yx,He:2006dk,Ma:2006sk,King:2006np,Morisi:2007ft,Bazzocchi:2007na,Bazzocchi:2007au,Lavoura:2007dw,Brahmachari:2008fn,Altarelli:2008bg,Bazzocchi:2008rz,Carr:2007qw,Feruglio:2007uu}by
the relations $\mathbf{1}^{0}\equiv\mathbf{1}$,
$\mathbf{1}^{+}\equiv\mathbf{1}'$,
$\mathbf{1}^{-}\equiv\mathbf{1}''$ and similarly for the doublet
representations. The representations $\mathbf{1}'$ and
$\mathbf{1}''$ are complex conjugated to each other, and the same
for the $\mathbf{2}'$ and $\mathbf{2}''$ representations. Since $T'$
is generated by the elements $S$ and $T$, we only need explicit
matrix representations of both $S$ and $T$ as follows
\begin{eqnarray}
\nonumber&&S(\mathbf{1}^{0})=S(\mathbf{1}^{+})=S(\mathbf{1}^{-})=1\\
\nonumber&&T(\mathbf{1}^{0})=1\;,~~~T(\mathbf{1}^{+})=\omega\;,~~~T(\mathbf{1}^{-})=\omega^2\\
\nonumber&&S(\mathbf{2}^{0})=S(\mathbf{2}^{+})=S(\mathbf{2}^{-})=N_1\\
\nonumber&&T(\mathbf{2}^{0})=\omega
N_2\;,~~~T(\mathbf{2}^{+})=\omega^2
N_2\;,~~~T(\mathbf{2}^{-})=N_2\\
\label{a3}&&S(\mathbf{3})=\frac{1}{3}\left(\begin{array}{ccc}
-1&2\,\omega&2\,\omega^2\\
2\,\omega^2&-1&2\,\omega\\
2\,\omega&2\,\omega^2&-1
\end{array}\right)\;,~~~T(\mathbf{3})=\left(\begin{array}{ccc}
1&0&0\\
0&\omega&0\\
0&0&\omega^2
\end{array}\right)
\end{eqnarray}
where $\omega=e^{i2\pi/3}$, and the matrices $N_1$ and $N_2$ are
defined as
\begin{equation}
\label{a4}N_1=\frac{-1}{\sqrt{3}}\left(\begin{array}{cc}
i&\sqrt{2}\,e^{i\pi/12}\\
-\sqrt{2}\,e^{-i\pi/12}&-i\\
\end{array}\right)\;,~~~N_2=\left(\begin{array}{cc}
\omega&0\\
0&1\\
\end{array}\right)
\end{equation}

\section{Higher order corrections to the vacuum alignment}
We will discuss how the vacuum alignment achieved at the leading
order is modified by the inclusion of higher dimensional operators,
then the superpotential $w_{v}$ is modified into
\begin{equation}
\label{b1}w_{v}=w^{LO}_v+w^{NL}_v
\end{equation}
where $w^{LO}_v$ is the leading order contributions
\begin{eqnarray}
\nonumber&&w^{LO}_{v}=g_1(\varphi^{R}_T\phi\phi)+g_2(\varphi^{R}_T\varphi_T)\Delta+g_3(\phi^R\phi)\chi+g_4(\varphi_T\phi^{R}\phi)+g_5\chi^R\chi^2+g_6\chi^R\theta'\theta''\\
\nonumber&&+g_7\chi^R(\varphi_T\varphi_T)+g_8\theta''^R\theta''^2+g_9\theta''^R\theta'\chi+g_{10}\theta''^R(\varphi_T\varphi_T)'+M_{\Delta}\Delta^R\Delta+g_{11}\Delta^{R}\chi^2\\
\nonumber&&+g_{12}\Delta^{R}\theta'\theta''+g_{13}\Delta^R(\varphi_T\varphi_T)+\bar{M}_{\Delta}\bar{\Delta}^R\bar{\Delta}+g_{14}\bar{\Delta}^{R}\Delta^2+g_{15}(\varphi^{R}_{S}\varphi_S\varphi_S)+g_{16}(\varphi^R_S\varphi_S)\tilde{\xi}\\
\label{b2}&&+g_{17}\xi^R(\varphi_S\varphi_S)+g_{18}\xi^R\xi^2+g_{19}\xi^R\xi\tilde{\xi}+g_{20}\xi^R\tilde{\xi}^2
\end{eqnarray}
The above leading order superpotential gives rise to the following
vacuum configuration
\begin{eqnarray}
\nonumber &&\langle \varphi_{T}\rangle=(v_{T},0,0),~~~\langle
\varphi_{S}\rangle=(v_S,v_S,v_S),~~~\langle\phi\rangle=(v_1,0),\\
\nonumber&&\langle\xi\rangle=u_{\xi},~~~\langle\tilde{\xi}\rangle=0,~~~\langle\theta'\rangle=u'_{\theta},~~~\langle\theta''\rangle=u''_{\theta},\\
\label{b3}&&\langle\Delta\rangle=u_{\Delta},~~~\langle\bar{\Delta}\rangle=\bar{u}_{\Delta},~~~\langle\chi\rangle=u_{\chi}
\end{eqnarray}
The effect of the next to leading order superpotential $w^{NL}_{v}$
on the above SUSY conserving vacuum configuration is just a shift in
the VEVs of the scalar fields, therefore the vacuum configuration is
modified into
\begin{eqnarray}
\nonumber&&\langle\varphi_{T}\rangle=(v_T+\delta v_{T1},\delta
v_{T2},\delta v_{T3}),~~~\langle\varphi_S\rangle=(v_S+\delta v_{S1},v_S+\delta v_{S2},v_S+\delta v_{S3}),\\
\nonumber&&\langle\phi\rangle=(v_1+\delta v_1,\delta
v_2),~~~\langle\xi\rangle=u_{\xi},~~~\langle\tilde{\xi}\rangle=\delta\tilde{u}_{\xi},~~~\langle\theta'\rangle=u'_{\theta}+\delta
u'_{\theta},\\
\label{b4}&&\langle\theta''\rangle=u''_{\theta}+\delta
u''_{\theta},~~~\langle\Delta\rangle=u_{\Delta}+\delta
u_{\Delta},~~~\langle\bar{\Delta}\rangle=\bar{u}_{\Delta}+\delta\bar{u}_{\Delta},~~~\langle\chi\rangle=u_{\chi}
\end{eqnarray}
and $w^{NL}_{v}$ is given by
\begin{eqnarray}
\nonumber w^{NL}_{v}&=&\frac{1}{\Lambda}\sum_{i=1}^{14}t_i{\cal
O}^{T}_i+\frac{1}{\Lambda^2}\big(f_1{\cal
O}^{\phi}_1+\sum_{i=1}^{8}k_i{\cal
O}^{\chi}_i+\sum_{i=1}^{4}c_i{\cal
O}^{\theta}_i+\sum_{i=1}^{8}d_i{\cal
O}^{\Delta}_i\big)+\frac{1}{\Lambda}\sum_{i=1}^{4}\bar{d}_i{\cal
O}^{\bar{\Delta}}_i
\end{eqnarray}
where ${\cal O}^{T}_i$, ${\cal O}^{\phi}_1$ etc are operators linear
in the driving fields $\varphi^{R}_T$ and $\phi^{R}$ et al., which
are consistent with the symmetry of the model, and each operator
comprises 4 or 5 superfields. Since the next to leading order
operators linear in $\varphi^{R}_{S}$ and $\xi^{R}$ are of order
$1/\Lambda^3$, therefore the shifts $\delta v_{Si}(i=1,2,3)$ and
$\delta\tilde{u}_{\xi}$ are suppressed by $1/\Lambda^3$, and these
operators are omitted in the $w^{NL}_v$ above. The operators ${\cal
O}^{T}_i(i=1-14)$ and ${\cal O}^{\phi}_1$  are given by
\begin{eqnarray}
\label{b5}&&\begin{array}{ll} {\cal
O}^{T}_1=(\varphi^{R}_T\varphi_T)(\varphi_T\varphi_T),&{\cal
O}^{T}_2=(\varphi^{R}_T\varphi_T)'(\varphi_T\varphi_T)'',\\
{\cal
O}^{T}_3=(\varphi^{R}_T\varphi_T)''(\varphi_T\varphi_T)',&{\cal
O}^{T}_4=((\varphi^{R}_T\varphi_T)_{\mathbf{3}_S}(\varphi_T\varphi_T)_{\mathbf{3}_S}),\\
{\cal
O}^{T}_5=((\varphi^{R}_T\varphi_T)_{\mathbf{3}_A}(\varphi_T\varphi_T)_{\mathbf{3}_S}),&{\cal
O}^{T}_6=(\varphi^{R}_T\varphi_T\varphi_T)\chi,\\
{\cal O}^{T}_7=(\varphi^{R}_T\varphi_T\varphi_T)''\theta',& {\cal
O}^{T}_8=(\varphi^{R}_{T}\varphi_T\varphi_{T})'\theta'',\\
{\cal O}^{T}_9=(\varphi^{R}_T\varphi_T)\chi^2,& {\cal
O}^{T}_{10}=(\varphi^{R}_T\varphi_T)\theta'\theta'',\\
{\cal O}^{T}_{11}=(\varphi^{R}_T\varphi_T)''\chi\theta',& {\cal
O}^{T}_{12}=(\varphi^{R}_T\varphi_T)''\theta''^2,\\
{\cal O}^{T}_{13}=(\varphi^{R}_T\varphi_T)'\chi\theta'',& {\cal
O}^{T}_{14}=(\varphi^{R}_T\varphi_T)'\theta'^2
\end{array}\\
\nonumber&&\\
\label{b6}&&{\cal O}^{\phi}_1=(\phi^{R}\phi)\Delta\bar{\Delta}^2
\end{eqnarray}

The structures ${\cal O}^{\chi}_i(i=1-8)$ are explicitly written as
follows
\begin{equation}
\label{b7}\begin{array}{ll} {\cal
O}^{\chi}_1=\chi^{R}\Delta\bar{\Delta}^2\chi,~~~~~~~~~~~~~~~~&{\cal
O}^{\chi}_2=\chi^{R}\Delta(\varphi_S\varphi_S\varphi_S),\\
{\cal O}^{\chi}_3=\chi^{R}\Delta(\varphi_S\varphi_S){\xi},&{\cal O}^{\chi}_{4}=\chi^{R}\Delta(\varphi_S\varphi_S)\tilde{\xi},\\
{\cal O}^{\chi}_5=\chi^{R}\Delta\xi^3,&{\cal
O}^{\chi}_6=\chi^{R}\Delta\xi^2\tilde{\xi},\\
{\cal O}^{\chi}_7=\chi^{R}\Delta\xi\tilde{\xi}^2,&{\cal
O}^{\chi}_8=\chi^{R}\Delta\tilde{\xi}^3
\end{array}
\end{equation}

The operators involving $\theta''^{R}$ and $\Delta^R$ are
\begin{eqnarray}
\label{b8}&&\begin{array}{ll} {\cal
O}^{\theta}_1=\theta''^{R}\theta'\Delta\bar{\Delta}^2,~~~~~~~~~~~~~~&{\cal
O}^{\theta}_2=\theta''^{R}\Delta(\varphi_S\varphi_S\varphi_S)',\\
{\cal
O}^{\theta}_3=\theta''^{R}\Delta(\varphi_S\varphi_S)'\xi,&{\cal
O}^{\theta}_4=\theta''^{R}\Delta(\varphi_S\varphi_S)'\tilde{\xi}
\end{array}\\
\nonumber&&\\
\label{b9}&&\begin{array}{ll} {\cal
O}^{\Delta}_1=\Delta^{R}\Delta\bar{\Delta}^2\chi,~~~~~~~~~~~~~~&{\cal
O}^{\Delta}_2=\Delta^R\Delta(\varphi_S\varphi_S\varphi_S),\\
{\cal O}^{\Delta}_3=\Delta^{R}\Delta(\varphi_S\varphi_S)\xi,&{\cal
O}^{\Delta}_4=\Delta^{R}\Delta(\varphi_S\varphi_S)\tilde{\xi},\\
{\cal O}^{\Delta}_5=\Delta^{R}\Delta\xi^3,&{\cal
O}^{\Delta}_{6}=\Delta^{R}\Delta\xi^2\tilde{\xi},\\
{\cal O}^{\Delta}_7=\Delta^{R}\Delta\xi\tilde{\xi}^2,&{\cal
O}^{\Delta}_{8}=\Delta^{R}\Delta\tilde{\xi}^3
\end{array}
\end{eqnarray}
and the operators ${\cal O}^{\bar{\Delta}}_i (i=1-4)$ are given by
\begin{eqnarray}
\label{b10}\begin{array}{ll} ~~~~~~~{\cal
O}^{\bar{\Delta}}_1=\bar{\Delta}^{R}\Delta(\varphi_T\varphi_T),~~~~~~~~~~~~~~~&{\cal O}^{\bar{\Delta}}_2=\bar{\Delta}^{R}\Delta\chi^2,\\
~~~~~~~{\cal
O}^{\bar{\Delta}}_3=\bar{\Delta}^{R}\Delta\theta'\theta'',&{\cal
O}^{\bar{\Delta}}_4=\bar{\Delta}^{R}(\varphi_T\phi\phi),\\
\end{array}
\end{eqnarray}
We perform the same minimization procedure as that in section V,
again we search for the zero of the $F$ terms associated with the
driving fields, Only terms linear in the shift $\delta v$ are kept,
and terms of order $\delta v/\Lambda$ are neglected, then the
minimization equations become
\begin{eqnarray}
\nonumber&&2ig_1v_1\delta v_1+g_2u_{\Delta}\delta
v_{T1}+g_2v_{T}\delta
u_{\Delta}+\frac{v_T}{\Lambda}(t_1v^2_T+\frac{4}{9}t_4v^2_T+\frac{2}{3}t_6u_{\chi}v_{T}+t_{9}u^2_{\chi}+t_{10}u'_{\theta}u''_{\theta})=0\\
\nonumber&&(1-i)g_1v_1\delta v_2+g_2u_{\Delta}\delta
v_{T3}+\frac{v_{T}}{\Lambda}(\frac{2}{3}t_{8}u''_{\theta}v_T+t_{13}u''_{\theta}u_{\chi}+t_{14}u'^2_{\theta})=0\\
\nonumber&&g_2u_{\Delta}\delta
v_{T2}+\frac{v_T}{\Lambda}(\frac{2}{3}t_7u'_{\theta}v_T+t_{11}u'_{\theta}u_{\chi}+t_{12}u''^2_{\theta})=0\\
\nonumber&&(g_3u_{\chi}+g_4v_{T})\delta v_2-(1-i)g_4v_1\delta
v_{T3}=0\\
\nonumber&&g_4v_1\delta v_{T1}-\frac{f_1}{\Lambda^2}u_{\Delta}\bar{u}^2_{\Delta}v_1=0\\
\nonumber&&g_{6}u'_{\theta}\delta u''_{\theta}+g_6
u''_{\theta}\delta u'_{\theta}+2g_{7}v_{T}\delta
v_{T1}+\frac{u_{\Delta}}{\Lambda^2}(k_1\bar{u}^2_{\Delta}u_{\chi}+3k_3u_{\xi}v^2_S+k_5u^3_{\xi})=0\\
\nonumber&&2g_{8}u''_{\theta}\delta u''_{\theta}+g_9u_{\chi}\delta
u'_{\theta}+2g_{10}v_T\delta
v_{T2}+\frac{u_{\Delta}}{\Lambda^2}(c_1u'_{\theta}\bar{u}^2_{\Delta}+3c_3u_{\xi}v^2_S)=0\\
\nonumber&&M_{\Delta}\delta u_{\Delta}+g_{12}u'_{\theta}\delta
u''_{\theta}+g_{12}u''_{\theta}\delta u'_{\theta}+2g_{13}v_{T}\delta
v_{T1}+\frac{u_{\Delta}}{\Lambda^2}(d_1\bar{u}^2_{\Delta}u_{\chi}+3d_3u_{\xi}v^2_S+d_5u^3_{\xi})=0\\
\label{b11}&&\bar{M}_{\Delta}\delta\bar{u}_{\Delta}+2g_{14}u_{\Delta}\delta
u_{\Delta}+\frac{1}{\Lambda}(\bar{d}_1u_{\Delta}v^2_T+\bar{d}_2u_{\Delta}u^2_{\chi}+\bar{d}_3u_{\Delta}u'_{\theta}u''_{\theta}+i\bar{d}_4v^2_1v_T)=0
\end{eqnarray}
Solving the above linear equations, then the shifts of the VEVs are
\begin{eqnarray}
\nonumber\delta
v_{T1}&=&\frac{f_1}{g_4}\frac{u_{\Delta}\bar{u}^2_{\Delta}}{\Lambda^2}\\
\nonumber\delta
v_{T2}&=&-\frac{v_T}{g_2u_{\Delta}\Lambda}(\frac{2}{3}t_7u'_{\theta}v_T+t_{11}u'_{\theta}u_{\chi}+t_{12}u''^2_{\theta})\\
\nonumber\delta
v_{T3}&=&-\frac{v_T}{2g_{2}u_{\Delta}\Lambda}(\frac{2}{3}t_8u''_{\theta}v_T+t_{13}u''_{\theta}u_{\chi}+t_{14}u'^2_{\theta})\\
\nonumber\delta
v_1&=&\frac{iv_T}{2g_1v_1\Lambda}(t_1v^2_T+\frac{4}{9}t_4v^2_T+\frac{2}{3}t_6u_{\chi}v_T+t_9u^2_{\chi}+t_{10}u'_{\theta}u''_{\theta})+{
\cal O}(\frac{1}{\Lambda^2})\\
\nonumber\delta
v_2&=&-\frac{(1-i)v_1}{4g_2u_{\Delta}\Lambda}(\frac{2}{3}t_8u''_{\theta}v_T+t_{13}u''_{\theta}u_{\chi}+t_{14}u'^2_{\theta})\\
\nonumber\delta
u'_{\theta}&=&-\frac{2g_{10}u'_{\theta}v^2_T}{3g_2g_8u''^2_{\theta}u_{\Delta}\Lambda}(\frac{2}{3}t_7u'_{\theta}v_T+t_{11}u'_{\theta}u_{\chi}+t_{12}u''^2_{\theta})+{
\cal O}(\frac{1}{\Lambda^2})\\
\nonumber\delta
u''_{\theta}&=&\frac{2g_{10}v^2_T}{3g_2g_8u''_{\theta}u_{\Delta}\Lambda}(\frac{2}{3}t_7u'_{\theta}v_T+t_{11}u'_{\theta}u_{\chi}+t_{12}u''^2_{\theta})+{\cal
O}(\frac{1}{\Lambda^2})\\
\nonumber\delta
u_{\Delta}&=&\frac{2(g_7g_{12}-g_6g_{13})f_1}{g_4g_6}\frac{u_{\Delta}\bar{u}^2_{\Delta}v_T}{M_{\Delta}\Lambda^2}+\frac{g_{12}u_{\Delta}}{g_6M_{\Delta}\Lambda^2}(k_1\bar{u}^2_{\Delta}u_{\chi}+3k_3u_{\xi}v^2_S+k_5u^3_{\xi})\\
\nonumber&&-\frac{u_{\Delta}}{M_{\Delta}\Lambda^2}(d_1\bar{u}^2_{\Delta}u_{\chi}+3d_3u_{\xi}v^2_S+d_5u^3_{\xi})\\
\label{b12}\delta\bar{u}_{\Delta}&=&-\frac{1}{\bar{M}_{\Delta}\Lambda}(\bar{d}_1u_{\Delta}v^2_T+\bar{d}_2u_{\Delta}u^2_{\chi}+\bar{d}_3u_{\Delta}u'_{\theta}u''_{\theta}+id_4v^2_1v_T)+{
\cal O}(\frac{1}{\Lambda^2})
\end{eqnarray}
where the contributions of order $1/\Lambda^2$ in $\delta v_1$,
$\delta u'_{\theta}$, $\delta u''_{\theta}$ and
$\delta\bar{u}_{\Delta}$, which are not written out explicitly, are
also higher order in $\lambda$ relative to the leading contributions
suppressed by $1/\Lambda$. Since $\delta v_{T1}$ and $\delta
u_{\Delta}$ are of order $1/\Lambda^2$, terms of the same order
should not be omitted in the relevant minimization equations, then
$\delta v_{T1}$ is modified into
\begin{eqnarray}
\nonumber\delta
v_{T1}&=&\frac{f_1}{g_4}\frac{u_{\Delta}\bar{u}^2_{\Delta}}{\Lambda^2}-\frac{v_T}{2g^2_2u^2_{\Delta}\Lambda^2}(\frac{2}{3}t_7u'_{\theta}v_T+t_{11}u'_{\theta}u_{\chi}+t_{12}u''^2_{\theta})(\frac{2}{3}t_8u''_{\theta}v_T+t_{13}u''_{\theta}u_{\chi}\\
\label{b13}&&+t_{14}u'^2_{\theta})
\end{eqnarray}
\end{appendix}


\end{document}